\documentstyle[amsfonts,amssymb,11pt,epsfig,diagram]{article}

\makeatletter
\@addtoreset{equation}{section}

\newtheorem{result}{Result}
\newtheorem{lemma}{Lemma}

\newcommand{\R}{{\Bbb R}}
\newcommand{\C}{{\Bbb C}}

\def\SL{{\rm SL}}
\def\SU{{\rm SU}}
\def\H{{{\bf H}^3}}
\def\Mob{\mbox{\rm M\"ob}}
\newcommand{\Sch}{{\mathfrak S}} 

\def\t{{\bf \theta}}
\def\A{{\bf A}}
\def\a{{\bf a}}

\newcommand{\be}{\begin{eqnarray}} 
\newcommand{\ee}{\end{eqnarray}}

\begin{document}

\begin{titlepage}

\thispagestyle{empty}

\title{On Holomorphic Factorization in\\
Asymptotically AdS 3D Gravity}

\author{
{\bf Kirill Krasnov}\thanks{{\tt 
krasnov@aei.mpg.de}}
\\ \\
{\it Physics Department, University of California, Santa Barbara, CA 93106, USA}\\
{\it and}\\
{\it Albert Einstein Institute, Golm/Potsdam, 14476, Germany}\\
{\it and}\\
{\it School of Mathematical Sciences, University of Nottingham}\\
{\it University Park, Nottingham, NG7 2RD, UK}}

\date{\normalsize February, 2003}
\maketitle

\begin{abstract}
\normalsize This paper studies aspects of ``holography'' for
Euclidean signature pure gravity on asymptotically AdS 3-manifolds. This theory
can be described as $\SL(2,\C)$ CS theory. However, not all configurations
of CS theory correspond to asymptotically AdS 3-manifolds. We show that configurations that do have 
the metric interpretation are parameterized by the so-called projective structures on the 
boundary. The corresponding asymptotic phase space is shown to be the cotangent bundle over the
Schottky space of the boundary. This singles out a ``gravitational'' sector of the $\SL(2,\C)$ CS theory. 
It is over this sector that the path integral has to be taken to obtain the gravity partition
function. We sketch an argument for holomorphic factorization of this
partition function.
\end{abstract} 

\end{titlepage}


\section{Introduction}
\label{sec:intr}

In this paper we study certain aspects of holography for negative cosmological constant 
gravity in 2+1 dimensions. The theory we consider is that of pure
gravity; the only field is the metric. This should be contrasted to the
by now standard setup of AdS${}_3$/CFT${}_2$ correspondence, in which the
3-dimensional theory contains, in addition to the metric  
(an infinite number of) other fields. Our main aim is to shed some 
light on ``holography'' in the pure gravity context. Namely, as was shown
more than ten years ago by Brown and Henneaux \cite{Brown-Hen}, the algebra of
asymptotic symmetries of negative cosmological constant 2+1 gravity
is the Virasoro algebra of certain central charge. Thus,
the corresponding quantum theory, if exists, must contain
the same algebra among its symmetries and is therefore a conformal
field theory. This argument of course applies not only to pure gravity, 
but also to any 3-dimensional theory containing it, in particular to the
system arising in the AdS${}_3$/CFT${}_2$ correspondence of string theory.
In that case the CFT is known: using the Maldacena limit argument 
\cite{Malda} one conjectures the CFT to be the sigma model describing the
low-energy dynamics of the D1/D5 system, see, e.g., the review
\cite{String-review} for more detail. There is no analogous
D-brane argument for the pure gravity case, so the question which
CFT, if any, gives a ``holographic'' description of pure gravity
cannot be answered this way. However, pure gravity is a topological
field theory. It has been known since the work of Witten \cite{Witten-J}
that 3d TQFT's are intimately related to 2d CFT's. One thus might 
suspect that some ``holographic'' description arises this way.
This paper is aimed at studying aspects of this ``holography''.

Some readers may object our usage of term ``holography'' to describe a
TQFT/CFT relation. Indeed, the bulk theory here has no propagating 
degrees of freedom. Holographic relations which are encountered in
string theory are, on the other hand, between a local theory with
propagating degrees of freedom in bulk and a local theory on the boundary. 
This is much more non-trivial than a TQFT/CFT relation. 
Still, in the TQFT/CFT context certain quantities of interest from the
CFT point of view can be very effectively calculated using the 
bulk theory, and vice versa. It is in this limited sense that
a TQFT/CFT relation is an example of holography.

Some may view this holography as trivial. However, as we shall
attempt to demonstrate in this paper, this is not so. First of all,
although gravity in 2+1 dimensions can be rewritten as a CS theory,
the relevant gauge group is {\it non-compact}. We are
thus entering a realm of non-compact TQFT's, which is much
less studied than the compact case. The problem here is
that, by analogy with the compact gauge group case, one expects 
quantum groups to be relevant, but now these are non-compact.
Even though non-compact quantum groups are studied to some extent, one
does not seem to understand them well enough to define
the corresponding non-compact TQFT's. The second point is that
in the usual well-understood compact CS TQFT/CFT correspondence
one has a relation only to a {\it holomorphic} sector of the CFT.
More precisely, the statement is that the Hilbert 
space $\cal H$ of the holomorphic conformal blocks of 
the group $G$ WZW CFT on a 
Riemann surface $X$ essentially coincides with the Hilbert space
of CS theory for group $G$ on a 3D manifold whose
boundary is $X$. In particular, the WZW holomorphic conformal blocks are
just certain states of the quantum CS theory. On the other hand,
the partition function of any CFT is constructed from both 
holomorphic and anti-holomorphic conformal blocks; one says that
it holomorphically factorizes. CS TQFT only gives one chiral sector.
To obtain the structure relevant for the full CFT, one 
needs two CS theories. The arising holographic correspondence
is rather non-trivial already in the compact gauge group case,
see \cite{Holography}. As it was shown in this work, given 
a ``chiral'' TQFT, e.g., CS theory with gauge group $G$, there
exists certain other TQFT, essentially given by two copies of
CS, such that the {\it full} CFT partition function receives
the interpretation of a state of this TQFT. For the compact
gauge group case this TQFT is given by the so-called Turaev-Viro model,
see \cite{Holography} for more detail. The present paper is a step
toward a non-compact version of that story. 

Thus, in the present paper we study the relation between a non-compact TQFT and the corresponding
holographic (full) CFT. The TQFT in question is Euclidean
3d gravity with negative cosmological constant. This 
Euclidean theory is interesting in its own right. Indeed, 
classically this is essentially the theory
of hyperbolic 3-manifolds --an extremely rich subject
that has been under active study for the last few decades. It
is a very interesting problem to construct the corresponding quantum
theory. This theory is expected to define new knot invariants, 
and may become of importance in 3D topology. The Euclidean quantum
theory also plays an important role in the construction \cite{String} of the Lorentzian
signature theory.

A relation between 3D gravity and a (full) CFT was anticipated
already in \cite{Verlinde}, and we were influenced by this
work when writing the present paper. The author notices
that the gravity action can be written as a difference of two
CS actions. At the level of the partition function this
suggests holomorphic factorization, a feature characteristic 
of a CFT. The author suggested that the relevant full CFT
is the quantum Liouville theory. Another work relevant in this regard
is \cite{Hen-Liouv}. This paper showed how the Liouville theory on
the boundary arises in asymptotically AdS gravity.

In the present paper we shall argue that the partition function of 3D gravity on
an asymptotically AdS manifold reduces to a full CFT partition function
on the boundary. Our argument is in the spirit of \cite{Verlinde} and is 
to show that the partition function holomorphically factorizes. However,
we are not claiming that the CFT in question is Liouville theory. In fact,
the quantum Liouville theory is known to be related to $\SL(2,\R)$ CS theory,
not $\SL(2,\C)$. Thus, the CFT that arises from 3-d gravity in the way
described in this paper is most probably not related to the Liouville theory.
It is some other CFT, whose exact nature is still to be determined.
We partially characterize this CFT by describing the relevant phase space.

The organization of this paper is as follows. We start by describing 
our main results and conclusions. Section \ref{sec:actions} 
gives the action formulation of the theory we are dealing with. We describe
the asymptotic phase space in section \ref{sec:phase-space}. The partition
function is studied in section \ref{sec:quantum}.

\section{Overview and main results}
\label{sec:results}

In this paper we are interested in the quantum theory
of negative cosmological constant gravity in three dimensions,
for the case of Euclidean signature. The action for this theory 
can be rewritten as the difference of two $\SL(2,\C)$ 
CS actions, see below. The corresponding CS quantum theory 
was studied in \cite{Witten-Complex}. However, the quantization 
procedure described there is not directly relevant in the 
{\it asymptotically} AdS case, because it does not
in any way incorporate the important asymptotic structure.
Work \cite{Witten-Complex} uses a parameterization of the $\SL(2,\C)$ CS phase
space as the cotangent bundle over the moduli space of
flat ${\rm SU}(2)$ connections. As we shall see below,
in the context of asymptotically AdS gravity certain other
parameterization is relevant. 

The $\SL(2,\C)$ WZW theory, or, more precisely, certain associated
coset theories were also actively studied in contexts other than 3D gravity. 
Thus, the gauged $\SL(2,\C)/{\rm SU}(2)$ WZW theory appears prominently 
in the usual ${\rm SU}(2)$ WZW or CS theories. Here
a problem of finding the scalar
product of CS states reduces to a problem of evaluating the 
$\SL(2,\C)/{\rm SU}(2)$ coset theory path integral, see
\cite{Gaw-case} and references therein. The coset theory
is rather well understood, in particular one knows explicitly 
the spectrum and the structure constants of the three point
function, see \cite{Tesch-coset}. Recently, some progress has also been
made in Liouville theory, which is a close relative of
the $\SL(2,\C)/{\rm SU}(2)$ coset model. Thus the work of
Ponsot and Teschner~\cite{Teschner} has proved
the Liouville bootstrap by reducing the problem to a 
question about representations of a certain non-compact
quantum group. Related is the development of quantum
Teichmuller spaces, see \cite{Fock} and \cite{Kashaev},
whose theory builds, essentially, upon representation
theory of the same quantum group. All these results are
potentially relevant for AdS 3D gravity.

On the physics side, AdS 3D gravity has been studied 
extensively, both classical aspects and the quantization.
Probably the most popular approach to the quantum theory 
is that based on the algebra of asymptotic symmetries,
see, e.g., \cite{Banados} and references therein. It 
originated in the paper by Brown and Henneaux \cite{Brown-Hen}. 
Studying the algebra of asymptotic symmetries of the
{\it Lorentzian} theory, they noticed that this algebra
essentially coincides with the Virasoro algebra of 
certain central charge. The central charge depends
on the cosmological constant, and was found to be equal to
$c=3l/2G$, where $l=1/\sqrt{-\Lambda}$ and $G$ is Newton's
constant. This means that a theory describing asymptotically 
AdS gravity must be a conformal field theory of this central 
charge. Coussaert, Henneaux and van Driel \cite{Hen-Liouv} then 
showed that the Einstein-Hilbert action reduces on shell, as a consequence
of asymptotically AdS boundary conditions, to the action of 
Liouville theory. This promoted Liouville
theory into a good candidate for the quantum theory of 
AdS 3D gravity. However, as we shall argue in this
paper, the actual holographic theory is {\it not} the Liouville,
but certain other, possibly related to it theory. 
The CS formulation was also used in tackling the
quantization problem. Thus, works \cite{Carlip,Banados-Ortiz} and more 
recently \cite{Kaul} showed that the BTZ BH entropy can be 
obtained from the ${\rm SU}(2)$ CS partition function 
by an analytic continuation.

Having completed this brief review, let
us outline the main constructions of the present paper. The paper
consists of two main parts. In the first part we  
analyze the structure of the space of classical solutions of the
theory. The main aim here will be to understand the asymptotic
structure of the CS connections. The second part is devoted to
an analysis of the gravity partition function. Here we will need the
results from the first part to state precisely over what class of fields 
the path integrals are taken.

The structure of the space of classical solutions is conveniently
summarized in a notion of {\it asymptotic phase space}. It is
introduced in section \ref{sec:phase-space}. This phase space is 
just the space of classical solutions of equations of motion, 
equipped with the symplectic structure that is
evaluated at the asymptotic infinity. 
For our purposes of analyzing the partition function we only need
to understand the structure of the space of solutions, more
precisely a certain parameterization of this space. The symplectic
structure on this space does not play any role, at least directly.
Indeed, what we consider in this paper is the partition function,
which is given by the path integral of $e^{-I_{\rm gr}}$. The 
symplectic structure on the space of solutions would be relevant
if we considered the path integral with an imaginary unit in 
the exponential, that is a quantum state. Such quantum states
would be essentially given by the quantization of our asymptotic
phase space. As an aside remark let us note that these states
play important role in the Lorentzian signature quantum 
theory, see \cite{String}. However, in the present paper we shall not use 
them. Thus, the
asymptotic phase space we get is not to be quantized, at least not 
in the present paper. The main rational for introducing it in the present 
paper is to point out that a natural symplectic structure
on the space of solutions is also the one induced by the gravity 
action.

As we explain in detail
in section \ref{sec:phase-space}, the spaces appearing as solutions
of the theory are handlebodies. The simplest example to keep in
mind is the AdS space itself, which, for Euclidean signature that
we are considering, is just the unit ball, its boundary being a sphere. 
For a more general space the conformal boundary at infinity is some 
Riemann surface, and the space itself is a handlebody. To understand
the structure of the space of solutions, let us recall that, since
there are no local DOF in 3D gravity, all constant negative curvature
spaces look locally like AdS. Thus, they can all be obtained from AdS
by discrete identifications. It is a standard result that
the moduli space of such manifolds with a Riemann surface $X$ as the 
boundary is parametrized by homomorphisms from $\pi_1(X)$ into 
the isometry group, which in this case is $\SL(2,\C)$, 
modulo the overall action of $\SL(2,\C)$. Our 
first result is that homomorphisms that arise 
in {\it asymptotically} AdS context are of a special type. Namely,
they are those associated with the so-called {\it projective
structures} on $X$. Thus, the space of solutions of our theory
is parametrized by moduli of the boundary $X$ and a projective
structure on $X$. It is known that this space is naturally 
a bundle over the moduli space, namely the cotangent bundle
$T^* T_g$, where $T_g$ is the Teichmuller space at genus $g$.
Kawai \cite{Kawai} has shown that the symplectic structure on
this space arising from its embedding into 
${\rm Hom}(\pi_1(X),\SL(2,\C))/\SL(2,\C)$ coincides with the usual cotangent 
bundle symplectic structure on $T^* T_g$. The first of this
symplectic structures is essentially that of CS theory, and thus 
also the gravitational one. Thus, the gravitational symplectic structure
evaluated at the asymptotic infinity coincides with the one on $T^* T_g$. 
Actually, as we shall see below, the phase space that appears is
the cotangent bundle over the so-called Schottky space. This is related to
the fact that the boundary in our approach is uniformized by the Schottky,
not Fuchsian groups. The Schottky space is a certain quotient of the
Teichmuller space. Summarizing, we get:
\begin{result} 
The asymptotic phase space of Euclidean
AdS 3D gravity is the cotangent bundle over the Schottky space of the boundary.
\end{result}

It is interesting to note that the same phase space is known to 
appear in 3D gravity in a different setting. For zero cosmological
constant, in the usual Lorentzian signature case, it
is a well-known result \cite{Moncrief} that the reduced phase space,
for spacetimes of topology $X\times\R$, where $X$ is some Riemann
surface, is the cotangent bundle over the Teichmuller space of 
$X$. To arrive to this result one uses, see \cite{Moncrief}, 
the usual geometrodynamics description, and a special 
time-slicing by hypersurfaces of constant York time $T={\rm Tr} K$, 
where $K$ is the hypersurface extrinsic curvature.
Witten \cite{Witten:2+1} arrived at the same result using the
Chern-Simons formulation. Let us note that the holomorphic factorization
of the partition function of a 3D theory is related to the fact that 
its phase space is $T^* T_g$. Indeed, the cotangent bundle 
$T^* T_g$ can be naturally identified with the space $T_g\times T_g$.
Therefore, quantum states of such a theory, which are
square integrable functions on the Teichmuller space, can be
also realized as $|\Psi|^2$ of states $\Psi$ obtained by
quantization of the Teichmuller space. Thus, interestingly,
in spite of the fact that the 
action of zero cosmological constant 3D gravity is not two CS actions, 
and the Verlinde \cite{Verlinde} argument for holomorphic 
factorization does not apply, the theory can still be expected to 
exhibit some analog of this property.

In section \ref{sec:quantum} we turn to an analysis of the partition
function. It is given by the path integral of 
$e^{-I_{\rm gr}[g]}$. Our point of departure is a representation of the gravitational partition
function as a path integral over the CS connections. As
we show in section \ref{sec:actions}, the boundary terms of
the gravity action are exactly such that in the
CS formulation one gets the following action:
\begin{equation}\label{action}
I[\A,\bar{\A}] := -iI_{\rm CS}^-[\A]+iI_{\rm CS}^+[\bar{\A}] + 
2\int d^2z \,\, {\rm Tr} \A_{\bar{z}} \bar{\A}_z.
\end{equation}
Here $I_{\rm CS}^-[\A], I_{\rm CS}^+[\bar{\A}]$ are CS 
actions suitable for fixing $\A_{\bar{z}}, \bar{\A}_z$ correspondingly,
see (\ref{CS-}), (\ref{CS+}). The key point is that
the real gravity action $I_{\rm gr}$ gets represented as 
$i I_{\rm CS}$ plus its complex conjugate. Thus, we consider
the partition function, which is the path integral of
the exponential of $(-I_{\rm gr})$. This path integral is not a quantum
state of our system precisely because there is no $i$
in the exponential. However, it gets represented in the
CS formulation as a product of two CS path integrals with the imaginary 
unit in the exponential, or in other words, two CS quantum states.
This is clearly resemblant of the holomorphic factorization.

To further analyze the structure of the partition function we
need to specify over which class of connections the path integral
is taken. We show in section \ref{sec:phase-space} that CS
connections appearing as classical solutions of our theory
have the following asymptotic structure. They are pure gauge:
\begin{equation}
\A \sim ({\bf m}_{T^\mu}\,{\bf F}_\mu\,{\bf h}_\varphi\,{\bf r})^{-1} 
d({\bf m}_{T^\mu}\,{\bf F}_\mu\,{\bf h}_\varphi\,{\bf r}),
\qquad \bar{\A} \sim 
({\bf \bar{r}}\,{\bf \bar{h}}_\varphi\,{\bf \bar{F}}_{\bar{\mu}}\,{\bf \bar{m}}_{\bar{T}^\mu})
d({\bf \bar{r}}\,{\bf \bar{h}}_\varphi\,{\bf \bar{F}}_{\bar{\mu}}\,
{\bf \bar{m}}_{\bar{T}^\mu})^{-1}
\end{equation}
Here ${\bf m}_{T^\mu}, {\bf F}_\mu, {\bf h}_\varphi, {\bf r}$ and correspondingly for the
other connection are certain (multi-valued) matrix-valued functions on $X$, to
be given below. The matrices ${\bf F}_\mu$ and ${\bf h}_\varphi$
depend in a certain way on the Beltrami differential $\mu$ and
the Liouville field $\varphi$ correspondingly. Beltrami differential
$\mu$ parameterizes the conformal structure on the boundary $X$. This
is achieved by fixing a reference conformal structure. Let $z$ be a
complex coordinate on the reference surface such that $|dz|^2$ gives
the corresponding conformal structure. Then $|dz+\mu d\bar{z}|^2$
is a metric in a different conformal class. Because of the conformal
anomaly to be discussed below, everything depends not just on the
conformal class of the metric, but also on a representative in each class.
The Liouville field $\varphi$ parameterizes different representatives in
the same conformal class. A representative is given by the metric
$e^\varphi |dz+\mu d\bar{z}|^2$. The matrix 
${\bf m}_{T^\mu}$ depends in a special way on a {\it quadratic differential}
$T^\mu$ on $X^\mu$ that is related to a projective structure. The matrix
${\bf r}$ is constant on $X$ and only depends on the radial coordinate.

The dependence of the connections on the radial coordinate is such that
the action (\ref{action}) contains only the logarithmic divergence. There
are no terms in (\ref{action}) containing the area-type divergence. One
can take care of the logarithmic divergence simply by introducing new
connections $\a, \bar{\a}$, such that the original connections are
gauge transforms of the new ones:
\begin{equation}
\A = \a^{\bf r}, \qquad \bar{\A} = {}^{\bf \bar{r}}\a.
\end{equation}
The new connections can be restricted to the boundary. 
The action (\ref{action}) considered as a functional of the 
connections $\a, \bar{\a}$ is explicitly finite. It however contains
a conformal anomaly coming from the last term in (\ref{action}). We
will define the CS path integral as an integral over $\a, \bar{\a}$.

We first analyze the genus zero case and then make comments as to the
general situation. The path integral can be taken in two steps. 
One first integrates
over the bulk, keeping the connections on the boundary fixed. Both 
${\cal D}\a$ and ${\cal D}\bar{\a}$ are the 
usual CS path integrals. For both connections the result is the exponential 
of the WZW action:
\begin{eqnarray}\label{CS-path}
\int {\cal D}\a \,\, e^{-iI_{\rm CS}^-[\a]} = e^{-I_{\rm WZW}[{\bf g}]}, 
\qquad \a|_{\partial M} = {\bf g}^{-1} d{\bf g}, \qquad 
{\bf g} = {\bf m}_{T^\mu}\,{\bf F}_\mu\,{\bf h}_\varphi, \\
\int {\cal D}\bar{\a} \,\, e^{iI_{\rm CS}^+[\bar{\a}]} = 
e^{-I_{\rm WZW}[{\bf \bar{g}}]}, 
\qquad \bar{\a}|_{\partial M} = {\bf \bar{g}} d{\bf \bar{g}}^{-1},
\qquad 
{\bf \bar{g}} = 
{\bf \bar{h}}_\varphi\,{\bf \bar{F}}_{\bar{\mu}}\,{\bf \bar{m}}_{\bar{T}^\mu}.
\end{eqnarray}
The result of the bulk integration is thus exponential of a new action. 
An important step is to realize that the WZW action 
$I_{\rm WZW}[{\bf m}_{T^\mu}\,{\bf F}_\mu]$ is essentially the
Polyakov light-cone gauge action~\cite{Polyakov}. In other words,
we have:
\begin{equation}
I_{\rm WZW}[{\bf m}_{T^\mu}\,{\bf F}_\mu] = \int d^2z \,\,
T\mu - W[\mu], \qquad
I_{\rm WZW}[{\bf \bar{F}}_{\bar{\mu}}\,{\bf \bar{m}}_{\bar{T}^\mu}] =
\int d^2z \,\, \bar{T}\bar{\mu} - W[\bar{\mu}].
\end{equation}
Here $T$ is a certain quadratic differential obtained from $T^\mu$.
When $\a$ is a solutions of classical equations of motion, that is flat,
the quadratic differential $T$ satisfies an equation involving $\mu$.
For example, when $\mu=0$ (no deformation of the reference surface $X$)
$T$ must be holomorphic. The quantity $W[\mu]$ above is the Polyakov action. 
It is a known function of $\mu$, which satisfies 
$\partial W[\mu]/\partial\mu = {\cal S}(f^\mu,z)$, where 
$f^\mu$ is the quasi-conformal mapping for $\mu$ and ${\cal S}$ stands
for the Schwartzian derivative. Using all these facts one gets
an explicit expression for the result of the bulk path integral.

The next step is to integrate over the boundary data. 
The partition function we are interested in is a functional of 
a conformal structure on the surface, and also of a representative
in this conformal class. Thus, it is a function of the
Beltrami differential $\mu, \bar{\mu}$, and of the Liouville field $\varphi$. To get this function
one has to integrate over the quadratic differential $T$ on $X$. Since one should integrate 
over all field configurations, not just classical solutions,
there are no additional constraints (like holomorphicity) that $T$ 
has to satisfy. Thus, one finds that the partition function has
the following simple structure:
\begin{equation}\label{part}
Z_{\rm gr}[\varphi,\mu,\bar{\mu}] = 
\int {\cal D}T {\cal D}\bar{T} \,\,
e^{-\int d^2z \,\, T\mu-\int d^2z \,\, \bar{T}\bar{\mu}
+W[\mu]+ W[\bar{\mu}] + K[\varphi,T,\bar{T},\mu,\bar{\mu}]}.
\end{equation}
Here $K[\varphi,T,\bar{T},\mu,\bar{\mu}]$ is a certain
functional, given in section \ref{sec:quantum}. 
An important fact is that it is a quadratic polynomial in $T, \bar{T}$. 
The integral over $T, \bar{T}$ can thus be easily taken. Shifting
the integration variables, and absorbing the result of a Gaussian
integral into a definition of the measure, one gets:
\begin{result} The partition function at genus zero holomorphically
factorizes according to:
\begin{equation}
Z_{\rm gr}[\varphi,\mu,\bar{\mu}] = 
e^{S_{\rm L}[\varphi,\mu,\bar{\mu}] + K[\mu,\bar{\mu}]}
\left[e^{-W[\mu]}\,e^{-W[\bar{\mu}]}\right].
\end{equation}
\end{result}
Here $S_{\rm L}[\varphi,\mu,\bar{\mu}]$ is the Liouville action
in the background $|dz+\mu d\bar{z}|^2$. The quantity $K[\mu,\bar{\mu}]$
is a certain functional of the Beltrami differential. The above result is exactly what one expects
as a holomorphically factorized partition function at genus zero.
We comment on a higher genus case in section \ref{sec:quantum}.
We sketch an argument for holomorphic factorization similar to
that of Witten~\cite{Witten:hol}. The argument is to interpret
the partition function as a certain inner product.

\section{Actions}
\label{sec:actions}

We start by defining the action for the theory, both in the
geometrodynamics and the CS formulations.

\noindent{\bf Geometrodynamics}

On a manifold with boundary one usually uses 
the following action
\be
-{1\over 2} \int d^3x \sqrt{g} (R+2) - \int d^2x \sqrt{q} K.
\ee
We have put $8\pi G = l =1$. For asymptotically AdS spaces
this action diverges. One of the two types of divergences can
be canceled \cite{Vijay} by adding an ``area'' term. The action becomes:
\begin{equation}
-{1\over 2} \int d^3x \sqrt{g} (R+2) - \int d^2x \sqrt{q} (K-1).
\end{equation}
The boundary condition for which (\ref{EH-action}) gives a
well-defined variational principle is that the induced boundary
metric is held fixed. However, the boundary in
our case is not a true boundary; rather it is only the conformal boundary
of the space. Thus, what is really being kept fixed in the variational
principle is the conformal class of the metric at the asymptotic infinity.
The Euclidean path integral with these
boundary conditions gives the canonical ensemble 
partition function, in which the intensive
thermodynamical parameters (temperature etc.) are kept fixed.

It turns out, however, that from the point of view of the Chern-Simons
formulation that we shall review shortly, a certain other action
is more natural. Namely, instead of fixing the induced boundary metric,
it is more convenient to fix the spin connection. In this case
no trace of extrinsic curvature term needs to be added. However,
one still needs the area term to cancel the divergence. Thus,
the action that we are going to use is:
\be\label{EH-action}
I_{gr} = -{1\over 2} \int d^3x \sqrt{g} (R+2) - \int d^2x \sqrt{q}.
\ee
This action can be viewed as suitable for computing the
micro-canonical ensemble partition function, in which the
energy, etc. is kept fixed at the boundary.

\noindent{\bf The Chern-Simons formulation}

The CS formulation of AdS 3D gravity has been extensively discussed
in the literature, see, e.g., \cite{Banados}. 
In this formulation Euclidean AdS 3D gravity becomes
the $\SL(2,\C)$ CS theory. This group is not semi-simple
and thus there are two possible choices of the trace to
be used when writing the action. As was explained in
\cite{Witten:2+1}, the trace to be used to get 
gravity is as follows. Let $J^i$ be generators
of rotations: $[ J^i, J^j]=\epsilon^{ijk}J^k$, and 
$P^i$ be generators of boosts: $[P^i,P^j]=-\epsilon^{ijk}J^k,
[P^i,J^j]=\epsilon^{ijk}P^k$. The trace to be used is
such that ${\bf\rm Tr}(J^i P^j)\sim\delta^{ij}$ and the
trace of $J$ with $J$ and $P$ with $P$ is zero. It is
customary to choose $J^i=-i\sigma^i, P^i=\sigma^i$,
where $\sigma^i$ are the usual Pauli matrices. Then the
trace can be written as ${\bf\rm Tr} = - {1\over 2}{\rm Im}\,{\rm Tr}$,
where ${\rm Tr}$ is the usual matrix trace. On the other hand,
the imaginary part can be represented as the difference 
of the quantity and its complex conjugate. Thus, the action
can be written using the ordinary matrix trace at the
expense of having to subtract the complex conjugate
action. The complex conjugate action can be thought
of as the CS action of the complex conjugate connection.
Thus, one has to work with both the original and the
complex conjugate connections simultaneously. 

Let us describe this in more detail.
With our choice of conventions (spelled out in  
Appendix~\ref{app:CS}) the two matrix valued 
CS connections are given by:
\begin{equation}\label{cs-connections}
\A = {\bf w} + {i\over 2} {\bf e}, \qquad
\bar{\A} = {\bf w} - {i\over 2} {\bf e}.
\end{equation}
They are complex and with
our conventions $\bar{\A} = - (\A)^\dagger$, where
$\dagger$ denotes Hermitian conjugation.
The quantities ${\bf w}, {\bf e}$ are the matrix valued spin connection
and the frame field correspondingly, see Appendix~\ref{app:CS}
for more details. The ``bulk'' CS action for $\A$ is: 
\begin{equation}\label{CS-action}
\tilde{I}_{\rm CS}[\A] = {1\over 2} \int_M {\rm Tr}
\left( \A\wedge d\A + {2\over 3} \A\wedge\A\wedge\A \right).
\end{equation}
The CS coupling constant, which is usually present in front of the
action in the combination $k/4\pi$ was set to $k=2\pi$. This is
for consistency with our choice $8\pi G = l =1$. 
Using the decomposition (\ref{cs-connections}) of $\A, \bar{\A}$ into
$\bf w, e$ one gets:
\begin{equation}
 -i \tilde{I}_{\rm CS}[\A] + i \tilde{I}_{\rm CS}[\bar{\A}] =
\int_M {\rm Tr} \left( {\bf e}\wedge{\bf f(w)}- 
{1\over 12}{\bf e\wedge e\wedge e}
\right) + {1\over 2} \int_{\partial M} {\rm Tr} ({\bf e}\wedge{\bf w}).
\end{equation}
The bulk term here is the usual Palatini action. When connection 
$\bf w$ satisfies its equation of motion, requiring that it is 
the spin connection compatible with $\bf e$, the action reduces to:
\begin{equation}\label{CS-action-geom}
 -i \tilde{I}_{\rm CS}[\A] + i \tilde{I}_{\rm CS}[\bar{\A}] \to
-{1\over 2} \int d^3x\,\, \sqrt{g} (R+2) - 
{1\over 2} \int d^2x\,\, \sqrt{q} K.
\end{equation}
We note that the boundary term here, although different from the one
in (\ref{EH-action}), also regularizes the action in the sense that
the action is at most logarithmically divergent.

Since we want the CS formulation action to reduce on shell to the
action (\ref{EH-action}), we need some extra 
boundary terms. As is clear from (\ref{CS-action-geom}),
the following quantity must be added:
\begin{eqnarray*}
+{1\over 2}\int d^2x\,\, \sqrt{q} K - \int d^2x\,\, \sqrt{q}
\end{eqnarray*}
The first term here is
\begin{eqnarray*}
- {1\over 2} \int {\rm Tr}\, {\bf e}\wedge{\bf w} = -{1\over 2i}\int
{\rm Tr}\,\A\wedge\bar{\A} = - \int d^2z\,\,{\rm Tr} (\A_z\bar{\A}_{\bar{z}}-
\A_{\bar{z}}\bar{\A}_z).
\end{eqnarray*}
Here we have introduced $d^2z=dz\wedge d\bar{z}/2i$. The area
term can also be expressed in terms of the CS connections. We have:
\begin{eqnarray*}
\int d^2x\,\, \sqrt{q} = -\int d^2z\,\, {\rm Tr} ({\bf e}_z 
{\bf e}_{\bar{z}}) = \int d^2z\,\, {\rm Tr}(\A-\bar{\A})_z
(\A-\bar{\A})_{\bar{z}}.
\end{eqnarray*}
The two terms combine into:
\begin{eqnarray*}
- \int d^2z\,\, {\rm Tr}(\A_z \A_{\bar{z}} + \bar{\A}_z \bar{\A}_{\bar{z}}
- 2 \A_{\bar{z}} \bar{\A}_z).
\end{eqnarray*}
Adding this expression to the bulk CS actions one gets:
\begin{equation}\label{action-CS}
I[\A,\bar{\A}] = -i I^-_{\rm CS}[\A]+
i I^+_{\rm CS}[\bar{\A}]
+ 2\int d^2z\,\, {\rm Tr}\, \A_{\bar{z}} \bar{\A}_z.
\end{equation}
Here $I^{\pm}_{\rm CS}[\A]$ are the CS actions suitable for
fixing $\A_z,\A_{\bar{z}}$ on the boundary correspondingly,
see Appendix \ref{app:WZW}.
We find it non-trivial that the boundary terms of the
geometrodynamics action combine in the CS formulation
into two ``holomorphic'' CS actions, plus a term that mixes
them. This is certainly suggestive of the holomorphic factorization.

\section{The Asymptotic Phase Space}
\label{sec:phase-space}

The purpose of this section is to understand in detail the
structure of the space of classical solutions of our theory.
In particular, we will analyze the asymptotic structure
of the CS connections. Facts derived in this section will
be used in an essential way in section \ref{sec:quantum}, when we
discuss the gravity path integral.

We summarize all the facts we obtain in this section in 
a notion of the asymptotic phase space. As we have briefly
explained in the introduction, this is just the 
space of solutions of equations of motion equipped with
a natural symplectic structure that is induced by the
gravity action. The symplectic structure is evaluated at
the conformal boundary. The phase space we introduce
is a Euclidean AdS${}_3$ analog of the asymptotic phase space of 
4D gravity, see \cite{Ashtekar}. The motivation in 4D comes from
the idea of asymptotic quantization put forward by Ashtekar.
He proposed to isolate radiative degrees of freedom in 
exact general relativity and then use the usual symplectic methods
to quantize them. This was achieved by introducing an ``initial value'' 
formulation with certain free data at future and past null 
infinities, instead of the usual extrinsic curvature
and the metric on a spatial hypersurface. The
asymptotic free data are parametrized by a certain connection field. 
The phase space is then the space of certain
equivalence classes of connections at future and past null infinity, 
with a rather natural symplectic structure, see \cite{Ashtekar}.
This phase space can be quantized using the usual methods. 
Our phase space is similar, except for the fact that we
are working with Euclidean metrics. The phase space will be
similarly parametrized by certain data at infinity, and the
gravitational action induces a certain symplectic structure.
One could quantize this phase space, the resulting states
turn out to be the analytic continuations
of the states of Lorentzian signature theory, see \cite{String}.
However, the main object of the present paper is not 
a quantum state, but the partition function. The difference
is that while the first can be realized as the path integral
of $e^{iI_{\rm gr}}$, the later is the path integral with
no imaginary unit in the exponential. 

The spaces that appear as classical solutions of our theory 
are Euclidean constant negative curvature manifolds that are
asymptotically AdS.\footnote{%
There is also a very rich class of hyperbolic manifolds arising as
complements of links in $S^3$. These play a major role in 3D topology,
see, e.g., \cite{Thurston} for a review. We do not consider these
spaces here.} A precise definition of asymptotically
AdS spaces was given, for the case of 3D, in \cite{Brown-Hen}. 
A nice coordinate-free description valid in any dimension can be found in 
\cite{Ashtekar-Magnon}. Both works treat Lorentzian signature spacetimes,
but with appropriate modifications the definition can be used also
in our Euclidean context. We restrict our attention only
to spaces that have the asymptotic boundary consisting of a single 
component. The boundary is then a Riemann surface. In this paper,
for simplicity, we shall consider only the case of compact Riemann 
surfaces, that is, no punctures or conical singularities. Our analysis
can be generalized to include punctures (and conical singularities)
but we will not consider this in the present paper. Let us note
in passing the physical interpretation of 3D hyperbolic spaces
having a compact Riemann surface as the asymptotic boundary. 
As was argued in \cite{K-Riemann}, these spaces should be 
interpreted as Euclidean continuations of multiple black 
hole solutions of \cite{Brill}. A particular case of the boundary
being a torus is the usual Euclidean version of the BTZ black hole.

Let us see now what is the structure of the space of such 3D
hyperbolic manifolds. Since there are no
local DOF, different geometries are obtained as quotients of AdS${}_3$,
or the hyperbolic space $\H$, 
by a discrete subgroup of its isometry group, which is ${\rm SL}(2,\C)$.
Such spaces $M$ can be parametrized by homomorphisms $\phi: \pi_1(M)\to
{\rm SL}(2,\C)$, modulo the overall action of ${\rm SL}(2,\C)$. 
The image of $\pi_1(M)$ under $\phi$ is just the
discrete group that is used to obtain the space: $M=\H/\phi(\pi_1(M))$.
Since our spaces $M$ have the topology of a handlebody, so that some
of cycles on the boundary are contractible inside $M$, the fundamental
group of $M$ is smaller than that of $\partial M$. However, as we shall
see, it is natural to allow singularities inside $M$. Then the
fundamental group of $M$ coincides with that
of $X$: $\pi_1(M) = \pi_1(X)$. Thus, solutions of equations of motion
are parametrized by homomorphisms $\phi\in{\rm Hom}(\pi_1(X),{\rm SL}(2,\C))$
modulo conjugation. The space of such 
homomorphisms has a natural symplectic structure, discussed in
\cite{Goldman} and thus becomes a phase space. For $X$ being a compact 
genus $g$ Riemann surface the (complex) dimension of this space is $6g-6$.
The described phase space, namely the space
of homomorphisms $\phi\in{\rm Hom}(\pi_1(X),\SL(2,\C))/\SL(2,\C)$,
is also the reduced phase space in the CS description. Indeed,
as is discussed in, e.g., \cite{Witten:2+1}, the reduced phase space
of CS theory on $X\times\R$ is parametrized by homomorphisms of
$\pi_1(X)$ into the gauge group in question, 
in our case $\SL(2,\C)$, modulo conjugation. Since gravity
is $\SL(2,\C)$ CS theory, the natural symplectic structure 
on $\phi\in{\rm Hom}(\pi_1(X),\SL(2,\C))/\SL(2,\C)$
is also the one induced by the gravity action. 

So far we have in no way used the asymptotic structure. As we
shall see, the asymptotic boundary
conditions restrict the type of homomorphisms 
$\phi\in{\rm Hom}(\pi_1(X),{\rm SL}(2,\C))$ that can arise
at infinity. The allowed homomorphisms turn out to be those
associated with {\it projective structures} on $X$. This
restricts one to a special smaller space, which still has
the complex dimension $6g-6$. This space is parametrized by
sections of the cotangent bundle over the Schottky space
space $\Sch_g$. We remind the reader that the Schottky space is a 
quotient of the Teichmuller space $T_g$ with respect to some of the modular transformations, see more on
this below. Unlike the
Teichmuller space, the Schottky space is not simply connected. On the cotangent bundle to the
Schottky space the CS symplectic  structure is known to reduce to the canonical cotangent 
bundle symplectic structure, see \cite{Kawai}. Thus, using the
asymptotically AdS boundary conditions one obtains the
cotangent bundle over $\Sch_g$ as the phase space. 

The phase space  ${\rm Hom}(\pi_1(X),\SL(2,\C))/\SL(2,\C)$
of $\SL(2,\C)$ CS theory is known to contain rather
nasty singularities, see, e.g., \cite{Goldman}.
However, it is non-singular near the homomorphisms
that come from projective structures,
see \cite{Kawai} and references therein.
Thus, asymptotically AdS boundary conditions serve
as a regulator, throwing away singular parts of
the CS phase space. Thus, asymptotically AdS gravity is {\it different}
from the $\SL(2,\C)$ CS theory in that the phase space of the theory,
even though of the same dimension, is smaller. Only certain CS configurations
have the metric interpretation. To obtain a theory related to gravity one should
only quantize this sector of CS. This can be compared with other examples of
CS-gravity correspondence. Thus, a well studied example is that of positive
cosmological constant Euclidean signature theory. The relevant CS gauge group
in that case is ${\rm SO}(4)\sim \SU(2)\times\SU(2)$. However, in this
example it is impossible to make a restriction to 
those CS configurations that have a metric interpretation.
More precisely, what is considered as gauge in CS theory should not
be treated as gauge in gravity. Thus, the reduced phase spaces of two
theories are different, see, e.g., \cite{Mat} for a good discussion of this. 
Quantization of CS theory gives a theory that in no obvious way is related to quantum gravity. 
On the other hand, in our case there is a clear cut sector of $\SL(2,\C)$ CS
that has a gravitational interpretation. It consists of those points
$\phi\in{\rm Hom}(\pi_1(X),{\rm SL}(2,\C))$ in the CS phase space that
come from projective structures on the boundary. By quantizing this
sector of CS one should get a theory that is quantum gravity, unlike
the case of positive cosmological sector, in which it is not known
how to select a gravitational sector. We consider the description of the
gravitational sector of $\SL(2,\C)$ CS theory as one of the most important
results of this paper.

There is another natural phase 
space that is associated with asymptotically AdS 3D gravity.
As we discuss in the next subsection, there is a large class
of 3D spaces arising from the Schottky uniformization of Riemann
surfaces. These spaces are not the most general ones appearing
in asymptotically AdS 3D gravity. More precisely, the most general solution is allowed
to have singularities inside the space. The restriction to non-singular solutions gives
exactly the spaces obtained via Schottky uniformization. One can consider the
restriction of the phase space $T^* \Sch_g$ to this smaller space
of solutions. This smaller space is $\Sch_g$ itself, and it is a Lagrangian sub-manifold in
$T^* \Sch_g$. 

To understand why homomorphisms that arise in asymptotically
AdS gravity are restricted to those coming from projective structures
we need to describe in more detail how the spaces in question can
be obtained by identifications of points.

\subsection{Asymptotically AdS spaces via Schottky groups}
\label{sec:ident}

In this subsection we describe how a large class of asymptotically AdS
3D manifolds can be obtained using Schottky groups. The spaces described in 
this subsection are not the most general asymptotically
AdS manifolds, as we shall see. However, once they are
understood, the structure of the whole space of solutions will become
clear. The material presented here is widely known, see, 
e.g., \cite{Hyperb-Man}. 

We will mostly use the Poincare upper half-space model
for $\H$. In this model the metric on $\H$ is given by:
\begin{equation}\label{ads}
ds^2 = {1\over\xi^2}(d\xi^2+|dy|^2).
\end{equation}
We have put the radius of curvature $l=1$. The boundary of $\H$ is
at $\xi=0$, and is just the (extended) complex plane $\C$, $y$ is a complex coordinate 
on the boundary.

The isometry group of $\H$ is denoted by $\Mob(\H)$.
It can be identified with the group of linear fractional transformations
$\Mob={\rm PSL}(2,\C)=\SL(2,\C)/\{\pm I\}$. This is done by considering 
the action of $\Mob$ on the boundary $\C$ of $\H$. It acts by fractional
linear transformations on $y$, or, the same, by conformal transformations.
Any fractional linear transformation can be expressed as a composition
of an even number of inversions with respect to circles or lines in $\C$.
To construct an isometry of $\H$ that corresponds to a particular
fractional linear transformation one has to extend the corresponding circles 
or lines to half-spheres or half-planes in $\H$. The isometry of $\H$ is
then given by the same composition of inversions in these half-spheres
and half-planes. This is called a Poincare extension of an element of
$\Mob$ to an element of $\Mob(\H)$. The Poincare extension can be realized
explicitly using quaternions, see, e.g., \cite{Hyperb-Man}.

A large class of asymptotically AdS spaces whose boundary is a compact genus
$g$ Riemann surface can be obtained using the so-called
{\it Schottky} groups. A Schottky group $\Sigma$ is a group that is
freely (that is, no relations) generated by a finite number of strictly
loxodromic (that is, no elliptic, no parabolic) elements of $\Mob$. A Schottky
group is called {\it marked} if one chooses in it a set of generators
$L_1,\ldots,L_g$ satisfying no relations. It is easiest to 
understand the structure of $\H/\Sigma$ by considering the action of
$\Sigma$ on the boundary of $\H$. Let us denote the completion of the
set of fixed points of this action by $\Delta$, 
and the complement of $\Delta$ in
$\C$ by $\Omega$. $\Sigma$ acts on $\Omega$ properly discontinuously,
and $\Omega/\Sigma\sim X$, where $X$ is a genus $g$ Riemann surface.
This is easiest to understand by introducing a fundamental region for
the action of $\Sigma$. Recall that a fundamental region $D\in\Omega$ is
a region no two points inside of which are related by a transformation
from $\Sigma$ and such that any point in $\Omega$ is obtainable as
an image of a point in $D$. A fundamental region for $\Sigma$ can
be obtained by picking up a set of $g$ pairs of non-intersecting
circles (or, more generally, Jordan curves) $C_1,\ldots,C_g$ and $C_1',\ldots,C_g'$, such that
$C_i'=-L_i(C_i)$ (minus denotes the change of orientation), which all 
lie outside of each other. The region outside of these circles is
a fundamental region $D$. The surface $X$ can be obtained from $D$ by identifying
the circles forming its boundary pairwise. 
As it is not hard to see, 
the surface one gets by identifying the circles --boundaries of the 
fundamental region D-- is indeed a $g$-handled sphere.

The classical {\it retro-section} theorem due to Koebe, see, e.g.,
\cite{Ford}, states that all Riemann surfaces can be obtained this 
way. Note, however, that the 3D space one obtains depends not
only on the conformal structure of surface $X$ (its boundary),
but also on the Schottky data: a set of $g$ non-intersecting
curves on $X$. One obtains the Schottky uniformization of $X$
cutting it precisely along this set of curves. Thus, there is
not a single Schottky uniformization of a given Riemann surface 
$X$, but infinitely many of them. Similarly, there is not
a single 3D manifold corresponding to a given conformal structure
on the boundary, but infinitely many such 3D manifolds. These 3-manifolds can
be thought of as different fillings of the surface $X$. One specifies a 3-manifolds
by saying which set of boundary cycles is contractible inside $M$. Moreover,
as we shall see below, even when the conformal structure of $X$ is 
fixed and the Schottky data are chosen, there 
is still a whole family of asymptotically AdS 3D manifolds approaching
at infinity this surface $X$. As we explain below, this family 
is parametrized by a projective structure on $X$. However, these
general manifolds are singular inside, as we shall see.

\subsection{The Fefferman-Graham asymptotic expansion}
\label{sec:exp}

To understand why the 3D spaces obtained via Schottky 
uniformization are not most general asymptotically AdS
spaces we need few more facts.
It can be shown that asymptotically AdS boundary conditions imply 
that asymptotically the metric has the following simple form:
\begin{equation}\label{exp}
ds^2 = {d\rho^2\over \rho^2} + q_{ab}\; dx^a dx^b,
\end{equation}
where 
\begin{equation}
q_{ab} = {1\over \rho^2} q^{(0)}_{ab} + q^{(1)}_{ab} + \rho^2 
q^{(2)}_{ab} + \ldots
\end{equation}
One can use Einstein equations to show that the trace part of 
$q^{(1)}_{ab}$ is completely determined by $q^{(0)}_{ab}$. The 
traceless part, however, is {\it free}. Once this trace-free part 
is specified, all other terms in the expansion 
are determined, see~\cite{FG,Graham}. In 3D, the freedom in
the traceless part of $q^{(1)}_{ab}$, as was noticed by,
e.g., Banados \cite{Banados} is exactly that of choosing
a quadratic differential on the boundary. Holomorphic quadratic differentials
are in one-to-one correspondence with equivalence classes
of projective structures, see Appendix \ref{app:proj}. This
is in agreement with the anticipated result that a general 
solution is parametrized
by both a conformal structure and an
equivalence class of projective structures on $X$.

For our purpose of analyzing the partition function, and also
to prove that the asymptotic phase space is the cotangent
bundle over the Teichmuller space, we need
an explicit parameterization of the space of solutions. To obtain
it, we shall find an explicit expression for the metric on
a 3D space obtained via Schottky uniformization. It will then
become quite clear how to modify this metric to obtain the
most general asymptotically AdS 3D space.

\subsection{The Banados, Skenderis-Solodukhin and Rooman-Spindel metrics}
\label{sec:BRS}

The result of Banados~\cite{Banados} is that in 3D, for the
case of flat boundary metrics, the Fefferman-Graham expansion 
(\ref{exp}) stops at order $\rho^2$. 
This result was initially obtained for the case of flat
boundary (genus one), but was later proved in full generality 
by Skenderis and Solodukhin \cite{SS}. What these authors
obtained is exactly the most general solution
of asymptotically AdS gravity. It was later shown by Rooman and Spindel 
\cite{RS} how this most general solution can be obtained by a coordinate
transformation from the BTZ metric.

Instead of simply borrowing the most general asymptotically AdS metric
from \cite{SS,RS} we sketch another derivation of it, which makes
clear the relation between the spaces obtained via the
Schottky groups and the metric \cite{SS,RS}. 
We use essentially the same idea as in \cite{RS}, however, our 
derivation is much simpler, for we apply
a coordinate transformation to AdS space, and not to the
BTZ BH space as in \cite{RS}. 

The idea of our derivation is to find a coordinate system in
AdS that is compatible with identifications from the Schottky
group $\Sigma$. The condition of compatibility is
that the AdS metric when written in this new coordinates is invariant
under the transformations from the Schottky group $\Sigma$. This metric
then descends to a metric in the quotient space. Using the
same method as in \cite{RS}, one finds that such a coordinate system 
is given by:
\begin{eqnarray}\label{coord}
\xi &=& {\rho\, e^{-\varphi/2}\over 1+ 
{1\over 4}\rho^2 e^{-\varphi} |\varphi_w|^2}, \\
\nonumber
y &=& w + {\varphi_{\bar{w}}\over 2} 
{\rho^2 e^{-\varphi}\over 1+ {1\over 4}\rho^2 e^{-\varphi} |\varphi_w|^2}.
\end{eqnarray}
The key quantity in this expressions is the {\it canonical Liouville
field} $\varphi$. It is a (real) function of the complex coordinate
$w\in\Omega : \varphi=\varphi(w,\bar{w})$. It depends in a certain way
on a conformal structure on $X$ and on the Schottky data. 
The canonical field $\varphi$ satisfies the Liouville
equation on the Schottky domain 
$\Omega$ and has the following transformation property under
the action of $\Sigma$:
\begin{equation}\label{transform}
\varphi(L w) = \varphi(w)-\ln{|L'|^2}.
\end{equation}
The Liouville field can be constructed from the map between the
Schottky and Fuchsian uniformization domains, see \cite{TZ}
for more details. The field $\varphi$ has a property that its
stress-energy tensor $T^\varphi$ is equal to the Schwartzian
derivative of the map $J^{-1}:\Omega\to{\bf H}$, where $\Omega$ is
the domain of discontinuity of $\Sigma$ and $\bf H$ is the
hyperbolic plane uniformizing the Riemann surface $X$, see
\cite{TZ}.

The coordinates (\ref{coord}) are compatible with the identifications.
This follows from the fact that
AdS$_3$ metric (\ref{ads}), when written in coordinates $\rho,w,\bar{w}$
is invariant under transformations from $\Sigma$. Indeed,
metric (\ref{ads}) expressed in terms of the new
coordinates becomes:
\begin{eqnarray}\label{RS-metric}
ds^2 = {d\rho^2\over\rho^2} + {1\over \rho^2}\, e^\varphi dw d\bar{w}
+ {1\over 2} T^\varphi dw^2 + {1\over 2} \bar{T}^\varphi d\bar{w}^2 + 
R\, dw d\bar{w} \\
\nonumber
+{1\over 4}\rho^2 e^{-\varphi} 
(T^\varphi dw + R\, d\bar{w})(\bar{T}^\varphi d\bar{w} + 
R\, dw).
\end{eqnarray}
Here we have introduced:
\begin{equation}
T^\varphi = 
\varphi_{ww} - {1\over 2} \varphi_w^2, \qquad R = \varphi_{w\bar{w}}.
\end{equation}
The first quantity is just the stress-energy tensor of the Liouville
field $\varphi$, the second is related to the curvature scalar of
the 2D metric $e^\varphi |dw|^2$. Using the transformation property
(\ref{transform}) of $\varphi$ one can show that:
\begin{equation}
(T^\varphi\circ L)(L')^2 = T^\varphi, \qquad (R\circ L)\overline{L'} L' = R.
\end{equation}
This immediately implies that (\ref{RS-metric}) is invariant under
the transformations $w\to L\circ w$ for all generators $L_i$ of $\Sigma_g$. 

The metric we just wrote is of the same form as the most general one 
obtained by Skenderis and Solodukhin \cite{SS}. Indeed, it can be 
written as:
\begin{eqnarray}
ds^2={d\rho^2\over\rho^2} + {1\over\rho^2}
\left(1+{\rho^2\over2}g_{(2)}g_{(0)}^{-1}\right)g_{(0)}
\left(1+{\rho^2\over2}g_{(2)}g_{(0)}^{-1}\right), \\
\nonumber
{g_{(2)}}_{ij}={1\over2}\left(R{(0)}{g_{(0)}}_{ij}+T_{ij}\right).
\end{eqnarray}
The notations here are self-explanatory. The metric (\ref{RS-metric})
is also the same as the one obtained by Rooman and Spindel~\cite{RS}.
Note, however, that this metric was obtained in \cite{RS} 
by applying a similar coordinate transformation to the BTZ black hole 
metric, not to the AdS space.\footnote{%
The fact that we have obtained the same metric 
starting from AdS is not surprising. Indeed, BTZ metric itself
can be obtained from AdS by a coordinate transformation.}
Thus, the
coordinate transformation (\ref{coord}) which relates the AdS
metric and the one given by (\ref{RS-metric}) is new. It can
be used, for example, to determine the range of coordinates
$\rho, w$ in (\ref{RS-metric}), a problem discussed in \cite{SS}.

It is now not hard to see how to modify the metric (\ref{RS-metric})
to obtain the most general asymptotically AdS manifold. Indeed,
the quantity $T^\varphi$, which was obtained to be equal to the
Liouville stress-energy tensor, does not have to be related
to $\varphi$. The metric (\ref{RS-metric}) is an exact solution of Einstein
equations for any meromorphic function $T(w)$ added to 
$T^\varphi$. Note, however, that the metric is non-singular inside 
the 3-manifold only when one uses $T^\varphi$. Indeed, non-singular manifolds
are the ones coming from the Schottky uniformization. A non-singular inside
manifold must have exactly $g$ cycles on the boundary that are
contractible in it. This is exactly the property of Schottky manifolds.
When $T$ is arbitrary, the metric (\ref{RS-metric}) is still a solution
of Einstein equations near the boundary, but it does not glue
globally into a non-singular metric inside. The typical singularity one obtains
is a line of conical singularities inside.

It is thus clear how to obtain the most
general metric from (\ref{RS-metric}).
One has to consider a general holomorphic quadratic differential 
added to $T^\varphi$. We choose to parameterize this  
as $T^\varphi-T$, where $T^\varphi$ is the
stress-energy tensor of the Liouville field $\varphi$ and
$T$ is some holomorphic quadratic differential for our Schottky group $\Sigma$:
$T(L w)(L')^2=T(w), L\in\Sigma$.
The most general asymptotically AdS metric then becomes:
\begin{eqnarray}\label{gen-metric}
ds^2 = {d\rho^2\over\rho^2} + {1\over \rho^2}\, e^\varphi dw d\bar{w}
+ {1\over 2} (T^\varphi-T) dw^2 + 
{1\over 2} (\bar{T}^\varphi-\bar{T}) d\bar{w}^2 + 
R\, dw d\bar{w} \\
\nonumber
+{1\over 4}\rho^2 e^{-\varphi} 
((T^\varphi-T) dw + R\, d\bar{w})((\bar{T}^\varphi-\bar{T}) d\bar{w} + 
R\, dw).
\end{eqnarray}
Here $R=\varphi_{w\bar{w}}$, and $\varphi$ is the (unique) solution of 
Liouville equation that has transformation properties (\ref{transform}). 
The above metric is of the same form as the
most general asymptotically AdS metric obtained in \cite{SS}. 
Having this explicit expression for the metric one can 
calculate the corresponding
CS connections. This is done in Appendix \ref{app:CS}. Their
asymptotic behavior is analyzed in the next subsection.

\subsection{The asymptotic structure of the CS connections}
\label{sec:CS-conn}

The asymptotic form of the CS connections corresponding to the
metric (\ref{RS-metric}) is obtained in Appendix~\ref{app:CS}, 
formulas (\ref{app2:4}),
(\ref{app2:5}). As is explained in the previous subsections, to
get the connections corresponding to (\ref{gen-metric}) we
have to replace $T^\varphi$ by $T^\varphi-T$ in all
the formulas.

It is not hard to notice that the dependence
on $\rho$ is correctly reproduced by introducing certain
new connections independent of $\rho$. Namely, one can see that
\begin{equation}\label{tr-r}
\A=(\a)^{\bf r} = {\bf r}^{-1} \a {\bf r} + {\bf r}^{-1} d{\bf r},
\qquad
\bar{\A}={}^{\bf \bar{r}}(\bar{\a}) = {\bf \bar{r}} 
\bar{\a} {\bf \bar{r}}^{-1} + 
{\bf \bar{r}} d{\bf \bar{r}}^{-1},
\end{equation}
where
\begin{eqnarray}\label{2-1}
\a_w = \left(\begin{array}{cc}
-{1\over 4}\varphi_w & {1\over 2}\,e^{-\varphi/2} (T^\varphi-T) \\
 e^{\varphi/2} & {1\over 4}\varphi_w 
\end{array} \right), \quad
\a_{\bar{w}} = \left(\begin{array}{cc}
{1\over 4}\varphi_{\bar{w}} & {1\over 2}\,e^{-\varphi/2} R \\
0 &  -{1\over 4}\varphi_{\bar{w}}
\end{array} \right) \\ \label{2-2}
\bar{\a}_w = \left(\begin{array}{cc}
-{1\over 4}\varphi_w & 0 \\
{1\over 2}\,e^{-\varphi/2} R & {1\over 4}\varphi_w 
\end{array} \right), \quad
\bar{\a}_{\bar{w}} = \left(\begin{array}{cc}
{1\over 4}\varphi_{\bar{w}} &  e^{\varphi/2} \\
{1\over 2} e^{-\varphi/2} (\bar{T}^\varphi-\bar{T})  &  
-{1\over 4}\varphi_{\bar{w}}
\end{array} \right)
\end{eqnarray}
and no $\a_\rho, \bar{\a}_\rho$ components. The complex matrix $\bf r$ is
given by:
\begin{equation}\label{r}
{\bf r} = \left(\begin{array}{cc}
{1\over \sqrt{i \rho}} &0 \\
0 & \sqrt{i \rho} \end{array}\right),
\end{equation}
and $\bf \bar{r}$ is the complex conjugate of $\bf r$. It does not
matter which branch of the square root is chosen in (\ref{r}). Note
that the new connections $\a, \bar{\a}$ no longer satisfy the
relation $\bar{\a}$ is minus the hermitian conjugate of 
$\a$. This happened because we used a $\sqrt{i}$ in the 
matrix defining the gauge transformation. The new relation is
that $\bar{\a}$ is the hermitian conjugate of $\a$, and the
diagonal components change sign. We did not have to introduce
the factor of $\sqrt{i}$ in the gauge transformation parameter
(\ref{r}). This would result in having some cumbersome factors of 
$i$ in the connections $\a, \bar{\a}$. We find our choice
for $\a, \bar{\a}$ more convenient. The price one pays is 
a little more complicated relation between the two 
connections.

It is not hard to see that the $\a_{\bar{w}}, \bar{\a}_w$ 
components of the connections are pure gauge: 
\begin{equation}\label{tr-h}
\a_{\bar{w}}={\bf h_\varphi}^{-1}\partial_{\bar{w}}{\bf h_\varphi}, \qquad
\bar{\a}_w = ({\bf \bar{h}_\varphi})\partial_w ({\bf \bar{h}_\varphi})^{-1},
\end{equation}
where
\begin{equation}\label{h}
{\bf h_\varphi} = \left(\begin{array}{cc}
e^{\varphi/4} & {1\over 2}\,\varphi_w e^{-\varphi/4} \\
0 & e^{-\varphi/4} \end{array} \right),
\qquad
{\bf \bar{h}_\varphi} = \left(\begin{array}{cc}
e^{\varphi/4} & 0 \\
-{1\over 2}\,\varphi_{\bar{w}} e^{-\varphi/4} & 
e^{-\varphi/4} \end{array} \right).
\end{equation}
We are thus led to new connections, which we will denote by
${\bf\alpha}, \bar{\bf\alpha}$. The connections $\a, \bar{\a}$ are
then the gauge transforms of ${\bf\alpha}, \bar{\bf\alpha}$:
\begin{equation}\label{alpha-h}
\a = ({\bf\alpha})^{\bf h_\varphi}, \qquad 
\bar{\a} = {}^{\bf \bar{h_\varphi}}(\bar{\bf\alpha}),
\end{equation}
with
\begin{eqnarray}\label{alpha-grav}
{\bf\alpha} = \left(\begin{array}{cc}
0 & -{1\over 2}T \\
1 & 0 \end{array}\right) dw, \qquad 
\bar{\bf\alpha} = \left(\begin{array}{cc}
0 & 1 \\
-{1\over 2}\bar{T} & 0
\end{array}\right) d\bar{w}.
\end{eqnarray}

What we have found (\ref{alpha-grav}) as the asymptotic form of the 
connections are exactly the canonical connections
$\alpha, \bar{\alpha}$ in holomorphic and anti-holomorphic 
vector bundles $E, \bar{E}$ of rank 2 over $X=\Omega/\Sigma$, see  
Appendix \ref{app:proj}. 

\subsection{General parameterization of the connections}
\label{sec:param}

What we have found in the previous subsection, is that the
CS connections $\a, \bar{\a}$ that arise from the most
general classical solution given by  
(\ref{gen-metric}) are of the form (\ref{alpha-h}) with
the connections $\bf\alpha, \bf\bar{\alpha}$ being the
canonical connections in the holomorphic and anti-holomorphic
vector bundles $E, \bar{E}$. These connections can in turn
be represented as gauge transforms, see (\ref{alpha-m}).
Thus, we get:
\begin{equation}
\a=({\bf m}_{T} {\bf h_\varphi})^{-1}
d ({\bf m}_{T} {\bf h_\varphi}), \qquad
\bar{\a} = 
({\bf\bar{h}}_\varphi {\bf\bar{m}}_{\bar{T}})
d ({\bf\bar{h}}_\varphi {\bf\bar{m}}_{\bar{T}})^{-1}.
\end{equation}
In this subsection we introduce a somewhat more general parameterization
of $\a, \bar{\a}$.

Let us recall that the CS connections $\a, \bar{\a}$ depend on
the conformal structure of $X$, and this
dependence comes through $\varphi$ and $T$, because $\varphi$ is the
Liouville field for $\Sigma$, see (\ref{transform}), and 
$T$ is a holomorphic quadratic differential for $\Sigma$. 
They also depend on 
a projective structure $f$, on which $T$ depends. It is somewhat
inconvenient to have the dependence on the moduli enter through
$\varphi$. Instead, as is usual in CFT, let us introduce a reference
Riemann surface $X$, and then consider quasi-conformal deformations
of it, parametrized by a Beltrami differential $\mu$. One then
gets more general connections $\a, \bar{\a}$ that depend on 
the conformal structure of the reference surface $X$, on a 
Beltrami differential $\mu$,
and on a projective structure on the surface $X^\mu$. These 
new connections are essentially the ones considered, e.g.,
by Kawai \cite{Kawai}. One obtains them considering the
holomorphic vector bundle $E^\mu$ over the deformed Riemann surface
$X^\mu$. There is the usual connection (\ref{alpha}) in
$E^\mu$, with $T=T^\mu$ being a holomorphic quadratic differential
for the deformed Schottky group $\Sigma^\mu$. As is explained in
\cite{Kawai}, this connection can be pulled back to the bundle
$E$ over the reference surface $X$. In fact,
$E^\mu$ is isomorphic to $E$. The isomorphism
is described by a matrix-valued function ${\rm F}_\mu(w)$:
\begin{equation}
{\bf F}_\mu = \left(\begin{array}{cc}
(f^\mu_w)^{-1/2} & {d\over dw} (f^\mu_w)^{-1/2}  \\
0  & (f^\mu_w)^{1/2}
\end{array}\right).
\end{equation}
Here $f^\mu$ is a solution of the Beltrami equation 
$f^\mu_{\bar{w}}=\mu f^\mu_w$ on the Schottky domain.
The matrix ${\bf F}_\mu$ satisfies the intertwining property:
\begin{equation}
\eta_{L^\mu}(f^\mu(w)) = {\bf F}_\mu(L w) \eta_L (w) 
{\bf F}_\mu^{-1}(w).
\end{equation}
Pulling back the canonical connection $\bf \alpha$ from 
$E^\mu$ to $E$ we get a new {\it flat} connection:
\begin{equation}
{\bf\alpha}^\mu = {\bf F}_\mu^{-1} f^\mu{}^* {\bf\alpha} {\bf F}_\mu
+ {\bf F}_\mu^{-1} d {\bf F}_\mu = \left(\begin{array}{cc}
0 & -{1\over 2}T\\
1 & 0
\end{array}\right) dw + \left(\begin{array}{cc}
-{1\over 2} \mu_w & -{1\over 2}(T\mu+\mu_{ww})\\
\mu & {1\over 2} \mu_w 
\end{array}\right) d\bar{w}.
\end{equation}
Here 
\begin{equation}\label{new-T}
T=T^\mu(f^\mu(w))(f^\mu_w)^2+{\cal S}(f^\mu,w)
\end{equation}
is a quadratic differentials for $\Gamma$, which is
however no longer holomorphic in $\bf H$. In fact, when $T^\mu$ is
holomorphic, that is comes from a projective structure on
$X^\mu$, one has:
\begin{equation}\label{eq-T}
(\partial_{\bar{w}} - \mu\partial_w - 2\mu_w)T=\mu_{www}.
\end{equation}
Similarly, one gets the other connection by pulling back
$\bf\bar{\alpha}$ from $\bar{E}^\mu$ with the help
of a matrix-valued function 
\begin{equation}
{\bf \bar{F}}_\mu=\left(\begin{array}{cc}
(\bar{f}^\mu_{\bar{w}})^{-1/2} & 0 \\
- {d\over d\bar{w}}  (\bar{f}^\mu_{\bar{w}})^{-1/2}  & 
(\bar{f}^\mu_{\bar{w}})^{1/2}
\end{array}\right).
\end{equation}
One gets:
\begin{equation}
{\bf\bar{\alpha}}^\mu = {\bf \bar{F}}_\mu f^\mu{}^* {\bf\bar{\alpha}} 
\, {\bf \bar{F}}_\mu^{-1}
+ {\bf \bar{F}}_\mu  d {\bf \bar{F}}_\mu^{-1} = \left(\begin{array}{cc}
0 & 1\\
-{1\over 2}\bar{T} & 0
\end{array}\right) d\bar{w} + \left(\begin{array}{cc}
{1\over 2} \bar{\mu}_{\bar{w}} & \bar{\mu}\\
-{1\over 2}(\bar{T}\bar{\mu}+\bar{\mu}_{\bar{w}\bar{w}}) & 
- {1\over 2} \bar{\mu}_{\bar{w}}
\end{array}\right) dw.
\end{equation}

Thus, instead of having all the dependence on the moduli in the
Liouville field $\varphi$ and the quadratic differential $T$, 
we introduce the new connections
${\bf\alpha}^\mu, {\bf\bar{\alpha}}^\mu$ that explicitly
depend on the Beltrami differential $\mu$. These are both
flat connections, but they are connections on different
bundles, namely on $E, \bar{E}$. Thus,
one cannot simply use them in the action
(\ref{action}), in particular because the cross-term
would not be invariantly defined. However, there exists
yet another bundle, to which both of these connections
can be mapped by a gauge transformation. The transformed connections become
connections on the same bundle, and the cross-term in
the action is defined. These pulled back connections are
just the gauge transforms of ${\bf\alpha}^\mu, {\bf\bar{\alpha}}^\mu$
with matrices ${\bf h}_\varphi, {\bf\bar{h}}_\varphi$
(\ref{h}). Thus, we get our desired final parameterization of
the CS connections:
\begin{equation}\label{param-1}
{\bf\alpha}^\mu = ({\bf m}_{T^\mu}\,{\bf F}_\mu)^{-1} d({\bf m}_{T^\mu}\,{\bf F}_\mu),
\qquad {\bf\bar{\alpha}}^\mu = 
({\bf \bar{F}}_{\bar{\mu}}\,{\bf \bar{m}}_{\bar{T}^\mu})
d({\bf \bar{F}}_{\bar{\mu}}\,{\bf \bar{m}}_{\bar{T}^\mu})^{-1},
\end{equation}
and
\begin{equation}\label{param-2}
\a = ({\bf\alpha}^\mu)^{{\bf h}_\varphi}, \qquad 
\bar{\a} = {}^{{\bf \bar{h}}_\varphi}({\bf \bar{\alpha}}^\mu).
\end{equation}

As we have said, the connections $\a, \bar{\a}$ are connections
on the same bundle over $X$, which we will denote by $P$. The
corresponding factor of automorphy is:
\begin{equation}\label{M}
{\bf M}=\left(\begin{array}{cc}
\left({\gamma'\over\bar{\gamma}'}\right)^{1/4} & 0 \\
0 & \left({\gamma'\over\bar{\gamma}'}\right)^{- 1/4}
\end{array}\right).
\end{equation}
Thus, the connections transform as:
\begin{equation}
\gamma^* \a = {\bf M}^{-1}\a {\bf M} + {\bf M}d{\bf M},
\qquad 
\gamma^* \bar{\a} = {\bf M}^{-1}\bar{\a} {\bf M} + {\bf M}d{\bf M}.
\end{equation}
Let us also note the transformation properties of the matrices
${\bf m}_T, {\bf F}_\mu, {\bf h}_\varphi$. We have:
\begin{eqnarray*}
{\bf m}_T(L w) = \chi^*_L {\bf m}_T(w) \eta_L^{-1}(w),
&\qquad& {\bf \bar{m}}_{\bar{T}}(L w) = 
\bar{\eta}_L^{-1}(w) {\bf \bar{m}}_{\bar{T}}(w) \bar{\chi}^*_L, \\
\left({\bf m}_{T^\mu} {\bf F}_\mu \right)(L w) = 
\chi^*_{L^\mu} \left({\bf m}_{T^\mu} {\bf F}_\mu \right)(w) \eta_L^{-1}(w),
&\qquad& \left({\bf \bar{F}}_\mu {\bf \bar{m}}_{\bar{T}^\mu}\right)(L w)= 
\bar{\eta}_{L}^{-1}(w) 
\left({\bf \bar{F}}_\mu {\bf \bar{m}}_{\bar{T}^\mu}\right)(w)
\bar{\chi}^*_{L^\mu}, 
\end{eqnarray*}
and
\begin{equation}\label{trans-g}
{\bf g}(L w) = \chi^*_{L^\mu} {\bf g}(w) {\bf M}, \qquad
{\bf\bar{g}}(L w)={\bf M}^{-1} 
{\bf\bar{g}}(w) \bar{\chi}^*_{L^\mu},
\end{equation}
where we have introduced
\begin{equation}\label{g}
{\bf g}= {\bf m}_{T^\mu} {\bf F}_\mu {\bf h}_\varphi, \qquad
{\bf\bar{g}}= {\bf\bar{h}}_\varphi {\bf \bar{F}}_\mu {\bf \bar{m}}_{\bar{T}^\mu}.
\end{equation}

\section{The partition function}
\label{sec:quantum}

Having the parameterization (\ref{param-1}), (\ref{param-2}) at
our disposal, we are ready to study the partition function. It is
given by the path integral of $e^{-I_{\rm gr}}$ over metrics.
As we have shown in section \ref{sec:actions}, the geometrodynamics
action, when written in the CS formulation, becomes a simple sum
of two CS actions, plus a cross-term that mixes them, see (\ref{action-CS}).
Thus, the partition function can be represented as the 
CS path integral. The boundary data that are kept fixed are
the conformal structure parametrized by the Beltrami differential
$\mu$ and the Liouville field $\varphi$. As is not hard to show, the action 
diverges logarithmically for the connection field with the
asymptotic behavior (\ref{app2:4}), (\ref{app2:5}). There is no
other divergence. The area type divergence was already taken care of.
To make the integrals well-defined, we should subtract the
logarithmically divergent term. A nice way to do this is to replace the
connections $\A, \bar{\A}$ by the gauge transformed connections
$\a, \bar{\a}$, see (\ref{tr-r}). This takes care of the
divergence but introduces a conformal anomaly, as expected of
a CFT partition function. Thus, we define the partition function
as:
\begin{equation}Z_{\rm gr}[\varphi,\mu,\bar{\mu}]=
\int {\cal D}\a {\cal D}\bar{\a}\,\, 
e^{-i I^-_{\rm CS}[\a]+
i I^+_{\rm CS}[\bar{\a}]
- 2\int d^2w\,\, {\rm Tr}\, \a_{\bar{w}} \bar{\a}_w}.
\end{equation}

Let us first analyze the genus zero case.
The path integral can be taken in two steps. One first integrates
over the bulk, keeping the connections on the boundary fixed. Both 
${\cal D}\a$ and ${\cal D}\bar{\a}$ are the 
usual CS path integrals. For both connections the result is the exponential 
of the WZW action:
\begin{eqnarray}\label{CS-path-int}
\int {\cal D}\a \,\, e^{-iI_{\rm CS}^-[\a]} = e^{-I_{\rm WZW}[{\bf g}]}, 
\qquad \a|_{\partial M} = {\bf g}^{-1} d{\bf g},  \\
\int {\cal D}\bar{\a} \,\, e^{iI_{\rm CS}^+[\bar{\a}]} = 
e^{-I_{\rm WZW}[{\bf \bar{g}}]}, 
\qquad \bar{\a}|_{\partial M} = {\bf \bar{g}} d{\bf \bar{g}}^{-1}.
\end{eqnarray}
Here $\bf g, \bar{g}$ are the matrix-valued functions introduced
in (\ref{g}).
The result of the bulk integration is thus exponential of a new action,
which we shall denote by $-I[\varphi,\mu,\bar{\mu},T,\bar{T}]$. Using the
Wiegman-Polyakov identity (\ref{WP}), one finds that  
this action is given by:
\begin{equation}\label{action-b}
I[\varphi,\mu,\bar{\mu},T,\bar{T}]= I_{\rm WZW}[{\bf g}{\bf\bar{g}}].
\end{equation}

To analyze the structure of the partition function another
representation turns out to be more convenient. As is not hard to show,
\begin{equation}\label{action-b-gauged}
I[\varphi,\mu,\bar{\mu},T,\bar{T}]=I_{\rm WZW}[{\bf m}_{T^\mu}\,{\bf F}_\mu] +
I_{\rm WZW}[{\bf \bar{F}}_{\bar{\mu}}\,{\bf \bar{m}}_{\bar{T}^\mu}] +
I_{\rm WZW}[{\bf g}_\varphi,{\bf\bar{\alpha}}^\mu_w,{\bf\alpha}^\mu_{\bar{w}}].
\end{equation}
Here ${\bf g}_\varphi = {\bf h}_\varphi {\bf\bar{h}}_\varphi$ and 
\begin{equation}
I_{\rm WZW}[{\bf g},{\bf\alpha}_w,{\bf\alpha}_{\bar{w}}] =
I_{\rm WZW}[{\bf g}] + 2\int d^2w\,\,{\rm Tr}  \left( 
{\bf g} {\bf\alpha}_w {\bf g}^{-1}
{\bf\alpha}_{\bar{w}} + {\bf\alpha}_w {\bf g}^{-1} \partial_{\bar{w}}
{\bf g} + {\bf\alpha}_{\bar{w}} {\bf g}\partial_w {\bf g}^{-1}  \right)
\end{equation}
is the usual gauged WZW action. An important step is to realize 
that the action $I_{\rm WZW}[{\bf m}_{T^\mu}\,{\bf F}_\mu]$ is the
Polyakov light-cone gauge action \cite{Polyakov}. In other words,
we have:
\begin{equation}\label{action-P}
I_{\rm WZW}[{\bf m}_T\,{\bf F}_\mu] = \int d^2w \,\,
T\mu - W[\mu], \qquad
I_{\rm WZW}[{\bf \bar{F}}_{\bar{\mu}}\,{\bf \bar{m}}_{\bar{T}}] =
\int d^2w \,\, \bar{T}\bar{\mu} - W[\bar{\mu}].
\end{equation}
Here $T$ and $S$ are given by (\ref{new-T}),
and $W[\mu]$ is the Polyakov action:
\begin{equation}\label{W}
W[\mu] = - {1\over 2} \int d^2w\,\, {f^\mu_{ww}\over f^\mu_w} \mu_w.
\end{equation}
The functional $W[\mu]$ satisfies
$\partial W[\mu]/\partial\mu = {\cal S}(f^\mu,w)$. 

We can use the representation (\ref{action-b-gauged}) together with
(\ref{action-P}) to integrate over the boundary data. Since we
would like the resulting partition function to depend on the
Beltrami differential $\mu, \bar{\mu}$ and $\varphi$, one only has the quadratic differential
$T$ to integrate over. Thus, we have
\begin{equation}
Z_{\rm gr}[\varphi,\mu,\bar{\mu}] = 
\int {\cal D}T {\cal D}\bar{T} \,\,
e^{-I[\varphi,\mu,\bar{\mu},T,\bar{T}]}.
\end{equation}
Here the integral is taken over all quadratic differentials
$T$ for $\Gamma$. One does not impose any equations like
(\ref{eq-T}) on $T$. Let us introduce
\begin{equation}
K[\varphi,T,\bar{T},\mu,\bar{\mu}]=
I_{\rm WZW}[{\bf g}_\varphi,{\bf\bar{\alpha}}^\mu_w,{\bf\alpha}^\mu_{\bar{w}}].
\end{equation}
This function can be explicitly calculated. It is of the form:
\begin{equation}
K[\varphi,T,\bar{T},\mu,\bar{\mu}]=
\int d^2w\,\, {1\over 2} e^{-\varphi} |\mu|^2 T\bar{T} + T(\ldots)
+\bar{T}(\ldots) + \ldots,
\end{equation}
where $(\ldots)$ denote certain terms depending on $\varphi$ and $\mu$.
Using (\ref{action-P}), one gets:
\begin{equation}\label{part'}
Z_{\rm gr}[\varphi,\mu,\bar{\mu}] = 
\int {\cal D}T {\cal D}\bar{T} \,\,
e^{- \int d^2w \,\,T\mu-\int d^2w \,\, \bar{T}\bar{\mu}- 
K[\varphi,T,\bar{T},\mu,\bar{\mu}]}\left[
e^{-W[\mu]}e^{-W[\bar{\mu}]}\right].
\end{equation}
Thus, what one gets is exactly the structure expected at genus zero. The
expression in the square brackets is just the product of holomorphic and
anti-holomorphic conformal blocks. Indeed, in the genus zero case the
Hilbert space is one-dimensional, and the holomorphic conformal block
is given by $\Psi[\mu]=e^{-W[\mu]}$. It satisfies the conformal 
Ward identity:
\begin{equation}
(\partial_{\bar{w}}-\mu\partial_w-2\mu_w){\delta W\over\delta\mu}=
\mu_{www}.
\end{equation}
The prefactor is also the expected one having to do with the conformal
anomaly. Indeed, it is not hard to show that, for $\mu=0$, the 
prefactor is exactly the Liouville action for $\varphi$. When
$\mu=0$ the expression in the exponential becomes simply 
$-I_{\rm WZW}[{\bf g}_\varphi]$. This can be shown to be:
\begin{equation}\label{Liouv}
I_{\rm WZW}[{\bf g}_\varphi]=-{1\over 2}\int d^2w\,\,
\left(|\varphi_w|^2-e^{-\varphi}R^2\right).
\end{equation}
Although this is not the usual Liouville action, which has
$e^\varphi$ as the second term, (\ref{Liouv}) does lead to
the Liouville equation as its equation of motion. It is
thus a version of the Liouville action. Thus, the prefactor
in (\ref{part'}), at least for $\mu=0$ is exactly the 
expected $e^{S_{\rm L}[\varphi]}$. The prefactor that comes
from the integration over $T, \bar{T}$ is absorbed in the
definition of the measure. For general $\mu$ one
has to carry the integration over $T$. It can be done by
completing the square. One then obtains a certain function of $\varphi$
and $\mu$ as the prefactor. We know that the dependence on
$\varphi$ must be that of the Liouville action 
$S_{\rm L}[\varphi,\mu,\bar{\mu}]$ in the
background metric $|dw+\mu d\bar{w}|^2$. There is also
an additional term, which we shall denote as $K[\mu,\bar{\mu}]$,
that we will not attempt to calculate. 
Summarizing, in the genus zero case one gets the holomorphically
factorized partition function:
\begin{equation}\label{part''}
Z_{\rm gr}[\varphi,\mu,\bar{\mu}] = 
e^{S_{\rm L}[\varphi,\mu,\bar{\mu}]+K[\mu,\bar{\mu}]}
\left[e^{-W[\mu]}e^{-W[\bar{\mu}]}\right].
\end{equation}

Let us now turn to a more complicated higher genus case. The
first complication that arises in this case is that the
WZW actions arising as the result (\ref{CS-path-int}) of the
CS path integral are not well-defined. Indeed,
the functions $\bf g, \bar{g}$ on $\bf H$ do not descend to
matrix-valued functions on $X$ because of their complicated
transformation property (\ref{trans-g}). Thus, the WZW action
functional is not well-defined on such $\bf g, \bar{g}$.
However, the full action, that is the WZW action 
(\ref{action-b}) of the product $\bf g\bar{g}$ 
is well-defined, at least in some cases, because
\begin{equation}\label{trans-gbarg}
({\bf g\bar{g}})(L w) = \chi^*_{L^\mu}({\bf g\bar{g}})(w) 
\bar{\chi}^*_{L^\mu}.
\end{equation}
Here $\chi^*_{\gamma^\mu}, \bar{\chi}^*_{\gamma^\mu}$ are
{\it constant} matrices, which are the monodromy representations
of the holomorphic and anti-holomorphic projective structures on $X^\mu$,
see Appendix \ref{app:proj}. One can convince oneself that,
in the case $\chi^*_\gamma\in{\rm SU}(1,1), \forall\gamma$ so
that $\chi^*_\gamma \bar{\chi}^*_\gamma={\bf 1}$, the
WZW action (\ref{action-b}) is well-defined. Indeed,
one only has to worry about the non-local WZ term. To define it,
one extends $\bf G$ to a function ${\bf G}(t,w,\bar{w})$
such that ${\bf G}(1,w,\bar{w})={\bf g\bar{g}}$ and 
${\bf G}(0,w,\bar{w})={\bf 1}$. Then the boundary terms
$\partial_w(\ldots), \partial_{\bar{w}}(\ldots)$ that arise
in the WZ term cancel pairwise from the components of the boundary
related by $\gamma$. However, the above condition on
the monodromy matrices is too restrictive and we don't want 
to impose it. Instead, we have to assume that the WZW actions
$I_{\rm WZW}[{\bf g}], I_{\rm WZW}[{\bf\bar{g}}]$ can be separately
defined. It must be possible to define them using a procedure similar
to that used in \cite{AT} to define a higher genus analog of the
Polyakov action. As is shown in \cite{AT}, one can define it as
an integral of the same integrand as in (\ref{W}) over a fundamental 
region on $\bf H$ plus a certain set of boundary terms, which 
make the action independent of a choice of the fundamental region,
and make the variational principle well-defined.
We shall assume that a similar procedure can be used to 
define the WZW actions (\ref{action-P}). The dependence on $T$
in this action may not be that simple as in (\ref{action-P}).
In fact, one does expect $T$ to enter the boundary terms 
that are necessary to make the action well-defined. We
shall also assume that one can make sense of the WZW action
$I_{\rm WZW}[{\bf g}_\varphi]$ in (\ref{action-b-gauged}).
Indeed, as we saw, this is essentially the Liouville action
for $\varphi$, which also does make sense on higher genus
surfaces, see \cite{TZ}. 

To show holomorphic factorization in the higher genus case we, 
following Witten \cite{Witten:hol},
interpret (\ref{part}) as a certain inner product. Witten noticed
that the partition function $Z_{\rm WZW}[A]$ of the WZW model coupled 
to the gauge field $A$ can be represented as an inner product:
\begin{equation}\label{part-WZW}
Z_{\rm WZW}[A] = \int {\cal D}B \,\, |\chi[A,B]|^2 
e^{\int d^2w\,\, B_w B_{\bar{w}}}.
\end{equation}
The key is that $\chi(A,B)$ depends on the holomorphic
component of $A$ and anti-holomorphic component of $B$,
and thus can be interpreted as a special state in 
${\cal H}_{\rm CS}\otimes\overline{{\cal H}_{\rm CS}}$.
It can be decomposed over a basis of states in
${\cal H}_{\rm CS}$:
\begin{equation}
\chi(A,B) = \sum_{I\bar{J}} \chi^{I\bar{J}} 
\Psi_I[A]\Psi_{\bar{J}}[B].
\end{equation}
Integrating over $B$ in (\ref{part-WZW}) one gets a
holomorphically factorized partition function:
\begin{equation}
Z_{\rm WZW}[A] = \sum_{I\bar{J}} N^{I\bar{J}} \Psi_I[A]
\Psi_{\bar{J}}[A].
\end{equation}
Actually, for the gauged WZW model $\chi^2=\chi$ and 
$\chi^{I\bar{J}}$ is just the identity matrix. This means that
$N^{I\bar{J}}$ is the inverse of the matrix giving the 
inner product of CS states:
\begin{equation}
N^{I\bar{J}}= (N_{I\bar{J}})^{-1}, \qquad
N_{I\bar{J}} = \int {\cal D}A\,\, \Psi_I[A]
\Psi_{\bar{J}}[A]. 
\end{equation}
Integrating over $A$
one gets the partition function of the gauged WZW model,
which, as can be seen from the above argument, is the
dimension of the CS Hilbert space. One does not expect a
similar ``projective'' property of $\chi$ in 3D gravity.

Witten's argument can be applied in our case if one notices that
the partition function can similarly be represented as
an inner product:
\begin{equation}\label{part-1}
Z_{\rm gr}[\varphi,\mu,\bar{\mu}]=
e^{K[\varphi,\mu,\bar{\mu}]}\int {\cal D}T{\cal D}\overline{T}\,\,
|\chi[\mu,T]|^2 e^{-{1\over 2}\int d^2w\,\, e^{-\varphi}
T\overline{T}},
\end{equation}
where $\chi[\mu,{\cal T}]$ can be thought of as a state in 
${\cal H}_{\Sch_g}\otimes\overline{{\cal H}_{\Sch_g}}$. Here 
${\cal H}_{\Sch_g}$ is the Hilbert space obtained
by quantizing the Schottky space $\Sch_g$. This needs some explanation.
The first argument of $\chi[\mu,T]$ is the Beltrami differential for $\Sigma$,
which is a holomorphic coordinate on $\Sch_g$. Thus, 
as a function of $\mu$, $\chi[\mu,T]$ can be thought of as a state
in ${\cal H}_{\Sch_g}$. The second argument is a quadratic differential
$T$. To understand why it can also be thought of as a holomorphic
coordinate on $\Sch_g$, let us recall some elements of Teichmuller
theory. As is well-known, there are two different realizations
of the Teichmuller space $T_g=T(\Gamma)$, where $\Gamma$ is the Fuchsian group for the
reference surface $X$. In one realization $T(\Gamma)$ is the
space of all Fuchsian groups $\Gamma^\mu$ obtained by 
quasi-conformal deformations $f^\mu: f^\mu_{\bar{z}}=\mu f^\mu_z$.
In order for $\Gamma^\mu=f^\mu\circ\Gamma\circ f^\mu{}^{-1}$
to be a Fuchsian group, the Beltrami differential must satisfy
certain ``reality condition'', namely 
$\mu(z,\bar{z})=\overline{\mu(\bar{z},z)}$. In the second 
realization $T(\Gamma)$ is the space of all quasi-Fuchsian groups.
In this case $\mu(z,\bar{z})=0$ in the lower half-plane $\bar{\bf H}$.
In this second case there is the so-called Bers embedding of
$T(\Gamma)$ into the space of quadratic differentials for
$\Gamma$ holomorphic in the lower half-plane. It is given by
the Schwartzian derivative of $f^\mu$. Because $f^\mu$ is
holomorphic in $\bar{\bf H}$, so is its Schwartzian derivative,
thus defining a holomorphic quadratic differential in $\bar{\bf H}$.
Such quadratic differentials thus provide holomorphic
coordinates on $T(\Gamma)$; this is one of the possible definitions
of the complex structure on $T(\Gamma)$. For the case in hand,
we work with the Schottky groups $\Sigma$, and 
$T$ defined by (\ref{new-T}) is a 
quadratic differential for $\Sigma$. Still it can be thought of as providing
a complex coordinate on $\Sch_g$. 

Then similarly to the gauged WZW case, one can decompose:
\begin{equation}
\chi[\mu,T] = \sum_{IJ} \chi^{IJ} \Psi_I[\mu]
\Psi_J[T].
\end{equation}
Using this decomposition, one gets a holomorphically
factorized partition function in the form:
\begin{equation}
Z_{\rm gr}[\varphi,\mu,\bar{\mu}] = 
e^{K[\varphi,\mu,\bar{\mu}]}
\sum_{I\bar{J}} N^{I\bar{J}}\Psi_I[\mu]\overline{\Psi_J[\mu]}.
\end{equation}
The quantities $N^{I\bar{J}}$ can in principle depend on the
moduli of the reference surface $X$. In the case of CS theory it
is known that a basis of the conformal blocks $\Psi_I[A]$ can 
be chosen in such 
a way that $N^{I\bar{J}}$ are moduli independent. It would be important
to establish an analogous property in our case. Similarly to the
CS case, it would require constructing
a vector bundle over $\Sch_g$, whose fibers are isomorphic to ${\cal H}_{\Sch_g}$,
and constructing a projectively flat connection in this bundle. We leave
this to future research.

\bigskip
\noindent
{\large \bf Acknowledgments}

I would like to thank L.\ Freidel, G.\ Horowitz and 
N.\ Nekrasov for discussions.
I am grateful to A.\ Ashtekar, J.\ Baez and L.\ Freidel for
very helpful comments on an earlier version of the manuscript,
and to L.\ Takhtajan for illuminating correspondence. Special
thanks are to the authors of \cite{Blau} for pointing out
an important sign mistake in an earlier version of this paper.
Thanks to the Center for Gravitational Physics and Geometry for
hospitality during the time that this paper was completed.
The author was supported in part by the NSF grant PHY95-07065.

\appendix

\section{CS and WZW actions}
\label{app:WZW}

The bulk CS action is given by:
\begin{equation}\label{app3:1}
\tilde{I}_{\rm CS}[\A] = {1\over 2} \int_M {\rm Tr}
\left( \A\wedge d\A + {2\over 3} \A\wedge\A\wedge\A \right).
\end{equation}
Variation of the bulk CS action gives a term
proportional to the equation of motion plus a boundary term:
\be\label{app3:2}
\delta\left(\tilde{I}_{\rm CS}[\A]\right) =
\int_M {\rm Tr}\,\left(\delta\A\wedge {\bf F}(\A)\right) 
-{1\over 2} \int_{\partial M}
{\rm Tr}\,\A\wedge\delta\A.
\ee
To have a well-defined action principle on a manifold with boundary one 
must specify boundary conditions. The standard procedure is to fix a 
complex structure on the boundary and fix either the holomorphic
or anti-holomorphic part of the connection. We shall use the following
rule to switch between differential form and complex notation:
\be
\int_{\partial M} A\wedge B = \int_{\partial M} dw \wedge d\bar{w} 
\left( A_w B_{\bar{w}} - A_{\bar{w}} B_w \right) =
2i \int_{\partial M} d^2w  
\left( A_w B_{\bar{w}} - A_{\bar{w}} B_w \right ),
\ee
where we have introduced the area element
\be
d^2w = {dw \wedge d\bar{w}\over 2i}.
\ee
With this convention, the action suitable for boundary condition
$\A_w$ kept fixed is:
\begin{equation}\label{CS+}
I^{+}_{\rm CS}[\A] = \tilde{I}_{\rm CS}[\A] + i \int d^2w\,\, {\rm Tr}
\,\A_w \A_{\bar{w}}.
\end{equation}
The action for $\A_{\bar{w}}$ fixed is:
\begin{equation}\label{CS-}
I^{-}_{\rm CS}[\A] = \tilde{I}_{\rm CS}[\A] - i \int d^2w\,\, {\rm Tr}
\,\A_w \A_{\bar{w}}.
\end{equation}

Under a gauge transformation 
\be
\A \to \A^{\bf g} = {\bf g}^{-1} \A {\bf g} + {\bf g}^{-1} d {\bf g}
\ee
the bulk action transforms as
\be
\tilde{I}_{\rm CS}[\A^{\bf g}] = \tilde{I}_{\rm CS}[\A]+C({\bf g},\A),
\ee
with the co-cycle 
\be
C({\bf g},\A)= -{1\over 6} \int_M {\rm Tr} 
\left( {\bf g}^{-1} d{\bf g}\right)^3 + {1\over 2} \int_{\partial M} {\rm Tr}
\left( \A\wedge d{\bf g} {\bf g}^{-1}\right).
\ee
Behavior  of the chiral actions under gauge transformations is as follows:
\begin{eqnarray}
I^{+}_{\rm CS}[\A^{\bf g}] = I^{+}_{\rm CS}[\A] -
i I_{\rm WZW}^{+}[{\bf g}] + 2i \int d^2w\,\,{\rm Tr} \A_w 
\partial_{\bar{w}}{\bf g} {\bf g}^{-1}, \\
I^{-}_{\rm CS}[\A^{\bf g}] = I^{-}_{\rm CS}[\A] +
i I_{\rm WZW}^{-}[{\bf g}] - 2i \int d^2w\,\,{\rm Tr} \A_{\bar{w}} 
\partial_w {\bf g} {\bf g}^{-1}.
\end{eqnarray}
Here the two WZW actions are given by: 
\be\nonumber
I_{\rm WZW}^{\pm}[{\bf g}] = -{1\over 4} \int d^2 x\,{\rm Tr} \left(
{\bf g}^{-1}\partial^\mu {\bf g} {\bf g}^{-1} \partial_\nu {\bf g} \right)
\mp {i\over 6} \int {\rm Tr}\, ({\bf g}^{-1} d{\bf g})^3 = \\
- \int d^2 w \,{\rm Tr} \left({\bf g}^{-1}\partial^w {\bf g} {\bf g}^{-1} 
\partial_{\bar{w}} {\bf g} \right) \mp {i\over 6} \int {\rm Tr}\, 
({\bf g}^{-1} d{\bf g})^3.
\ee
The minus sign in front of the first term is standard. It 
makes the non-linear sigma-model action (first term) 
positive for $\bf g$ unitary.
The action $I_{\rm WZW}^+$ gives as its equation of motion that the
current $J_{\bar{w}}={\bf g}^{-1} \partial_{\bar{w}} {\bf g}$ is conserved 
$\partial_w J_{\bar{w}}=0$, while $I_{\rm WZW}^-$ implies that
$J_w={\bf g}^{-1} \partial_w {\bf g}$ is conserved.
Note that $I_{\rm WZW}^{-}[{\bf g}]=I_{\rm WZW}^{+}[{\bf g}^{-1}]$.

We shall also need the Polyakov-Wiegman identity:
\be\label{WP}
I_{\rm WZW}^+[{\bf gh}] = I_{\rm WZW}^+[{\bf g}]+ I_{\rm WZW}^+[{\bf h}]
- 2\int d^2w\,\, {\rm Tr}\, {\bf g}^{-1} \partial_w {\bf g} 
\partial_{\bar{w}} {\bf h} {\bf h}^{-1}, \\
I_{\rm WZW}^-[{\bf gh}] = I_{\rm WZW}^-[{\bf g}]+ I_{\rm WZW}^-[{\bf h}]
- 2\int d^2w\,\, {\rm Tr}\, {\bf g}^{-1} \partial_{\bar{w}} {\bf g} 
\partial_w {\bf h} {\bf h}^{-1}.
\ee

\section{Chern-Simons connections}
\label{app:CS}

In this appendix we obtain explicit expressions for the 
complex CS connections corresponding to the metric (\ref{RS-metric}).
Let us rewrite the metric (\ref{RS-metric})
in the complex frame form as:
\begin{equation}
ds^2 = \left( {d\rho\over\rho} \right)^2 +
\t_w \t_{\bar{w}},
\end{equation}
where
\begin{equation}
\t_w = {1\over 2}\rho\, e^{-\varphi/2} T^\varphi \,dw +
{1\over \rho}\,e^{\varphi/2} (1+{1\over 2}\rho^2\, e^{-\varphi} R) d\bar{w},
\end{equation}
and $\t_{\bar{w}} = \overline{\t_w}$. Let us also
introduce the real frame:
\begin{equation}
e_1 = d\rho/\rho, \qquad
e_2 = {1\over 2}(\t_w+\t_{\bar{w}}),\qquad 
e_3 = {1\over 2i} (\t_w - \t_{\bar{w}}).
\end{equation}
It is not hard to find the spin connection coefficients. One gets:
\begin{eqnarray}\nonumber
w_{12} = {1\over 2\rho}\,e^{\varphi/2}
\left[ \left(1-{1\over 2}\rho^2\,e^{-\varphi}(R+T^\varphi)\right)dw+
\left(1-{1\over 2}\rho^2\,e^{-\varphi}(R+\bar{T}^\varphi)
\right)d\bar{w} \right], \\
\label{app2:1}
w_{31} = {i\over 2\rho}\,e^{\varphi/2}
\left[ \left(1-{1\over 2}\rho^2\,e^{-\varphi}(R-\bar{T}^\varphi)
\right)d\bar{w}
-\left(1-{1\over 2}\rho^2\,e^{-\varphi}(R-T^\varphi)\right)dw \right], 
\\ \nonumber
w_{23} = {1\over 2i}(C\,dw - \bar{C} d\bar{w}).
\end{eqnarray}
Here the quantity $C$ is given by a rather complicated
expression:
\begin{eqnarray}\nonumber
&{}&C = {1\over 
(1+ {1\over 2}\rho^2\,e^{-\varphi} R)^2 - \rho^2 e^{-2\varphi} 
T^\varphi \bar{T}^\varphi}
\Big[ \varphi_w + \rho^2\,e^{-\varphi} (R_w-T^\varphi_{\bar{w}}) \\
&{}& +{1\over 4}\rho^4\,e^{-2\varphi} (2T^\varphi\bar{T}^\varphi_w + 
2RT^\varphi \varphi_{\bar{w}} + 
2RR_w -
T^\varphi \bar{T}^\varphi \varphi_w - R^2 \varphi_w - 
2RT^\varphi_{\bar{w}} - 2T^\varphi R_{\bar{w}})
\Big].
\end{eqnarray}
Here we gave the full expression without using the fact that
$T^\varphi,\bar{T}^\varphi$ are conserved. Note that in the limit $\rho\to 0$
\begin{equation}
C=\varphi_w.
\end{equation}
We also note that in the flat case $R=0$, and using the conservation
laws for $T^\varphi,\bar{T}^\varphi$ one gets $C=\varphi_w$ exactly and 
not just in the $\rho\to 0$ limit.

Let us now find the CS connections. It is customary to represent these
in the matrix form. We introduce the anti-hermitian matrices
$J^1 = i\sigma^3, J^2=i\sigma^2, J^3=i\sigma^1$, 
where $\sigma^i$ are the standard Pauli matrices. Then define:
\begin{equation}
{\bf e} = e_i J^i = \left(\begin{array}{cc}
id\rho/\rho & \t_w \\
-\t_{\bar{w}} & -id\rho/\rho 
\end{array}\right).
\end{equation}
The metric is then given by:
\begin{equation}
ds^2 = -{1\over 2} {\rm Tr}({\bf e}{\bf e}).
\end{equation}
The CS connections are defined as:
\begin{equation}
A_i = w_i + i\,e_i, \qquad \bar{A} = w_i - i\,e_i,
\end{equation}
where 
\begin{equation}
w_i = - {1\over2}\epsilon_{ijk} w_{jk}.
\end{equation}
Let us also introduce the matrix valued connections $\A,\bar{\A}$:
\begin{equation}
\A = {1\over 2} A_i J^i, \qquad \bar{\A} = {1\over 2} \bar{A}_i J^i,
\end{equation}
or, equivalently
\begin{equation}
\A = {\bf w} + {i\over 2} {\bf e}, \qquad
\bar{\A} = {\bf w} - {i\over 2} {\bf e},
\end{equation}
where ${\bf w} = (1/2) w_i J^i$. The factor of $1/2$ in the decomposition
of the connections into $J^i$ is adjusted so that
the curvatures of $\A,\bar{\A}$ are given by the usual expressions
${\bf F}(\A) = d\A + \A\wedge\A$ and similarly for $\bar{\A}$.
Using the expressions (\ref{app2:1}) for the spin connection we
get:
\begin{eqnarray}\label{app2:2}
\A = \left(\begin{array}{cc}
-{d\rho\over 2\rho} - {1\over 4}C dw + {1\over 4}\bar{C} d\bar{w}&
{i\over 2}\rho\,e^{-\varphi/2}(T^\varphi dw + Rd\bar{w}) \\
-{i\over \rho}\,e^{\varphi/2} dw &
+{d\rho\over 2\rho} + {1\over 4}C dw - {1\over 4}\bar{C} d\bar{w}  
\end{array}\right) \\ \label{app2:3}
\bar{\A} = \left(\begin{array}{cc}
+{d\rho\over 2\rho} - {1\over 4}C dw + {1\over 4}\bar{C} d\bar{w}&
-{i\over \rho}\,e^{\varphi/2} d\bar{w} \\ 
{i\over 2}\rho\,e^{-\varphi/2}(R dw + \bar{T}^\varphi d\bar{w}) &
-{d\rho\over 2\rho} + {1\over 4}C dw - {1\over 4}\bar{C} d\bar{w}   
\end{array}\right)
\end{eqnarray}
Note that
\begin{equation}
\bar{\A} = - (\A)^\dagger.
\end{equation}
The above expressions for the CS connections are also found in
\cite{RS-1}. In the genus one case the boundary geometry is flat
$R=0$ and these expressions essentially coincide with the ones found,
e.g., by Banados \cite{Banados}. To see this, we note that in the flat
case the Liouville field is given by the sum of holomorphic and
anti-holomorphic pieces $\varphi=A+\bar{A}$, and $e^{A/2}$ can be
absorbed into the complex coordinate $w$. This removes the
exponentials and the factors proportional to $C$ on the 
diagonal and the resulting connections are exactly those of
\cite{Banados} (up to some sign differences which stem from
a difference in conventions). 

We will also need the expressions for the connections $\A,\bar{\A}$
in the limit $\rho\to 0$. These are obtained by replacing $C$
in (\ref{app2:2}), (\ref{app2:3}) with $\varphi_w$. One gets, in
components:
\begin{equation}\label{app2:4}
\A_\rho \sim \left(\begin{array}{cc}
-{1\over 2\rho} & 0 \\
0 & {1\over 2\rho} \end{array}\right), \quad
\A_w \sim \left(\begin{array}{cc}
-{1\over 4}\varphi_w & {i\over 2}\rho\,e^{-\varphi/2} T^\varphi \\
-{i\over \rho} \, e^{\varphi/2} & {1\over 4}\varphi_w 
\end{array} \right), \quad
\A_{\bar{w}} \sim \left(\begin{array}{cc}
{1\over 4}\varphi_{\bar{w}} & {i\over 2}\rho\,e^{-\varphi/2} R \\
0 &  -{1\over 4}\varphi_{\bar{w}}
\end{array} \right)
\end{equation}
and 
\begin{equation}\label{app2:5}
\bar{\A}_\rho \sim \left(\begin{array}{cc}
{1\over 2\rho} & 0 \\
0 & - {1\over 2\rho} \end{array}\right), \quad
\bar{\A}_w \sim \left(\begin{array}{cc}
-{1\over 4}\varphi_w & 0 \\
{i\over 2}\rho\,e^{-\varphi/2} R  & {1\over 4}\varphi_w 
\end{array} \right), \quad
\bar{\A}_{\bar{w}} \sim \left(\begin{array}{cc}
{1\over 4}\varphi_{\bar{w}} & -{i\over \rho} \, e^{\varphi/2}  \\
{i\over 2}\rho\,e^{-\varphi/2} \bar{T}^\varphi &  -{1\over 4}\varphi_{\bar{w}}
\end{array} \right)
\end{equation}

\section{Projective structures}
\label{app:proj}

There is a one-to-one correspondence between holomorphic 
quadratic differentials on a Riemann surface and equivalence 
classes of {\it projective structures}. 
Thus, the DOF described by $T, \bar{T}$ are those of a 
projective structure. Let us review this correspondence.
We use \cite{Gunning} as the main source. We give a description in 
terms of Fuchsian groups and Fuchsian uniformization. Analogous
facts hold in the Schottky picture, except that one uses the
whole complex plane instead of the hyperbolic plane ${\bf H}$.

A projective structure on a Riemann surface $X={\bf H}/\Gamma$
is a complex analytic function $f(z)$ on the covering space $H$
that satisfies:
\begin{equation}\label{proj-structure}
f(\gamma\circ z) = \chi_\gamma \circ f(z), \qquad \forall\gamma\in\Gamma.
\end{equation}
Here $\chi_\gamma$ is a representation of $\gamma\in\Gamma$ in
the group $\Mob$ that acts on $f(z)$ by a fractional linear
transformation. Thus, projective structures on $X$ are in
one-to-one correspondence with representations 
$\chi\in{\rm Hom}(\Gamma,\Mob)$. Inequivalent projective structures
are defined as corresponding to inequivalent representations,
where equivalent representations are those related by a conjugation
in $\Mob$. 

Here are some examples of projective structures. The simplest
example is that of the {\it Fuchsian} projective structure.
In this case the function $f$ is the identity: $f(z)=z$,
and the representation $\chi$ of $\Gamma$ is the Fuchsian
group itself. The second example is the {\it Schottky}
projective structure. Having a Schottky uniformization
of a given Riemann surface $X$ (it depends on a choice of
the maximal set of non-intersecting, homotopy non-trivial
and non-equivalent set of curves on $X$), there is a 
map $J(z)$ from the unit disc $H$ to the complex plane of the Schottky 
uniformization:
\begin{equation}\label{diagram-1}
\begin{diagram}
\node{H} \arrow[2]{e,t}{J} \arrow{se,b}{\pi_\Gamma}
\node[2]{\Omega} \arrow{sw,r}{\pi_\Sigma} \\
\node[2]{X}
\end{diagram}
\end{equation}
Here $\pi_\Sigma$ is the quotient map $\pi_\Sigma:\Omega\to\Omega/\Sigma=X$,
where $\Sigma$ is the corresponding Schottky group.
The map $J(z)$ gives the Schottky projective structure:
\begin{equation}
J(A\circ z) = J(z), \qquad J(B\circ z) = L\circ J(z).
\end{equation}
Here $A\in\Gamma$ correspond to the elements of $\pi_1(X)$
that go around the handles along which one cuts to
obtain the Schottky uniformization, and $L\in\SL(2,\C)$
are the generators of the Schottky group.

Let us now discuss a one-to-one correspondence between equivalence classes of 
projective structures and holomorphic quad\-ratic differentials for 
$\Gamma$. The relation is that
the Schwartzian derivative ${\cal S}(f)$ of the function $f(z)$ defining 
the projective structure gives a holomorphic quadratic differential:
${\cal S}(f)\circ\gamma\, (\gamma')^2 = S(f)$. 
The opposite is also true. Namely,
given a holomorphic quadratic differential $T$, there exists a 
solution $f$ of the Schwartz equation
\begin{equation}
{\cal S}(f;z) = T.
\end{equation}
The solution is unique modulo the natural action of $\Mob$ on the left. 
Thus, a quadratic
differential defines a projective structure up to equivalence. All in all,
we have:
\begin{equation}
f(z) - {\rm proj.\,\, structure} \Longleftrightarrow
{\cal S}(f;z)\, dz^2 - 
{\rm holom.\,\, quadratic\,\, differential\,\, for\,\, \Gamma}
\end{equation}

There is a canonical lift of the representation 
$\xi\in{\rm Hom}(\Gamma,\Mob)$ to a representation in $\SL(2,\C)$.
It is given by the so-called {\it monodromy representation} $\chi^*$ of a 
projective structure. Namely, consider the so-called Fuchs equation:
\begin{equation}\label{Fuchs}
u'' + {1\over 2} T u = 0.
\end{equation}
Here $T$ is a holomorphic quadratic differential on $X={\bf H}/\Gamma$.
The monodromy group of this equation gives a representation $\chi^*$ of
$\Gamma$ in $\SL(2,\C)$. To obtain the monodromy
representation one uses the following simple lemma \cite{Gunn}:
\begin{lemma} If $u, v$ are two linearly independent solutions of
the Fuchs equation (\ref{Fuchs}), then
\begin{eqnarray*}
{1\over\sqrt{\gamma'}}
\left(\begin{array}{c} u \\ v \end{array}\right)(\gamma z)
\end{eqnarray*}
also satisfies the same equation with respect to $z$.
\end{lemma}
This immediately means that
\begin{equation}
{1\over\sqrt{\gamma'}}
\left(\begin{array}{c} u \\ v \end{array}\right)(\gamma z)=
\chi^*_\gamma
\left(\begin{array}{c} u \\ v \end{array}\right)(z),
\end{equation}
where $\chi^*_\gamma$ is some matrix independent of $z$.
This is the monodromy representation. 
The ratio of two linearly independent solutions
$f=u/v$ satisfies ${\cal S}(f;z)=T$ and gives a projective
structure. 

The last fact we need is a relation between projective structures and
equivalence classes of rank 2 holomorphic complex vector bundles over 
$X$. Given the monodromy representation 
$\chi^*\in{\rm Hom}(\Gamma,\SL(2,\C))$
of a projective structure, one gets a
holomorphic vector bundle over $X$ as a quotient of the trivial bundle
$\C^2\times {\bf H}$ over $\bf H$. The equivalence relation used
to get the quotient is:
$\C^2\times {\bf H} \ni \{F, z\} \sim \{\chi^*(\gamma) F, \gamma\circ z\}$.
Here $F$ is a vector $F=(u,v)$.

In practice it is more convenient to work with a somewhat different,
but related bundle over $X$. The following definitions and facts are
from \cite{Gunn,Kawai}. Let us introduce the following
holomorphic rank 2 vector bundle $E$ over $X$: $E=(z,F)/\sim$,
where the equivalence relation is: $(z,F)\sim(\gamma z, \eta_\gamma(z) F)$.
Here $\eta_\gamma(z)$ is a {\it factor of automorphy}, given by:
\begin{equation}\label{eta}
\eta_\gamma(z) = \left(\begin{array}{cc}
(\gamma')^{-1/2} &  {d\over dz} (\gamma')^{-1/2}\\
0 &  (\gamma')^{1/2} \end{array}\right).
\end{equation}
Consider the following holomorphic connection in $E$:
\begin{equation}\label{alpha}
{\bf\alpha} = \left(\begin{array}{cc}
0 & -{1\over 2} T \\
1 & 0 \end{array}\right) \, dz.
\end{equation}
Here $T$ is some holomorphic quadratic differential for $\Gamma$.
As is not hard to check, this connection can also be represented
as:
\begin{equation}\label{alpha-m}
{\bf\alpha} = {\bf m}^{-1} d{\bf m},
\end{equation}
where ${\bf m}$ is given by:
\begin{equation}\label{m}
{\bf m} = \left(\begin{array}{cc}
u &  u_z \\
v &  v_z \end{array}\right),
\end{equation}
and $u, v$ are two linearly independent solutions of
the Fuchs equation (\ref{Fuchs}) for $T$. The holomorphic
connection $\bf\alpha$ is a connection in the bundle $E$
for it satisfies the following transformation property:
\begin{equation}
\eta_\gamma^{-1} \gamma^* {\bf\alpha} \, \eta_\gamma +
\eta_\gamma^{-1} d \eta_\gamma = {\bf\alpha}.
\end{equation}
Here $\gamma^* \bf\alpha$ is the pullback of the connection
$\bf\alpha$ under the mapping $z\to\gamma z$. Let us finally
note one more lemma showing a relation between the holonomy of
the connection $\bf\alpha$ and the monodromy matrix
$\chi^*_\gamma$. 
\begin{lemma} Let matrix ${\bf m}(z)\in\SL(2,\C)$ be given by
(\ref{m}) with $u, v$ being two linearly independent solutions
of the Fuchs equation. Then
\begin{equation}
{\bf m}(\gamma z) \eta_\gamma(z) = \chi^*_\gamma {\bf m}(z).
\end{equation}
\end{lemma}
This shows that the holonomy of the connection $\bf\alpha$ along
some cycle on $X$ is
essentially (up to the factor of automorphy) the corresponding
monodromy matrix $\chi^*$.

As is discussed in~\cite{Gunning}, not
all classes of representations $\chi^*\in{\rm Hom}(\Gamma,\SL(2,\C))$
correspond to projective structures. The representations that are
excluded are the unitary and reducible ones, or the ones that become
unitary or reducible when restricted to any subgroup of finite index
in $\Gamma$.

To summarize, there is a set of relations between projective structures
$f$, holomorphic quadratic differentials $T={\cal S}(f)$, solutions $u, v:
u/v=f$ of the Fuchs equation (\ref{Fuchs}) for $T$, and holomorphic 
rank 2 vector bundles with the canonical holomorphic connection given by
(\ref{alpha}).

Let us now quickly state analogous facts for anti-holomorphic
projective structures. Equivalence classes of such are in one-to-one
correspondence with anti-holomorphic quadratic differentials
$\bar{T}$. One can similarly introduce the anti-holomorphic
Fuchs equation
\begin{equation}\label{Fuchs'}
\bar{u}''+ {1\over 2}\bar{T} \bar{u}=0.
\end{equation}
\setcounter{lemma}{0}
\renewcommand{\thelemma}{\arabic{lemma}'}
\begin{lemma}
If $\bar{u}, \bar{v}$ are two linearly independent solutions of
the Fuchs equation (\ref{Fuchs'}), then
\begin{eqnarray*}
{1\over\sqrt{\bar{\gamma}'}}
\left(\begin{array}{cc} \bar{u} & -\bar{v} \end{array}\right)(\gamma z)
\end{eqnarray*}
also satisfies the same equation with respect to $\bar{z}$.
\end{lemma}
This means that
\begin{equation}
{1\over\sqrt{\bar{\gamma}'}}
\left(\begin{array}{cc} \bar{u} & -\bar{v} \end{array}\right)(\gamma z)=
\left(\begin{array}{cc} \bar{u} & -\bar{v} \end{array}\right)(z)
\, \bar{\chi}^*_\gamma.
\end{equation}
Here $\bar{\chi}^*_\gamma$ is related to $\chi^*_\gamma$ as follows:
\begin{equation}
\chi^*_\gamma = \left(\begin{array}{cc}
a & b\\
c & d \end{array}\right), \qquad
\bar{\chi}^*_\gamma = \left(\begin{array}{cc}
\bar{a} & -\bar{c}\\
-\bar{b} & \bar{d} \end{array}\right).
\end{equation}

One can similarly introduce an 
anti-holomorphic vector bundle $\bar{E}$. It is defined using the
following factor of automorphy:
\begin{equation}
\bar{\eta}_\gamma(\bar{z}) = \left(\begin{array}{cc}
(\bar{\gamma}')^{-1/2} & 0 \\
- {d\over d\bar{z}}(\bar{\gamma}')^{-1/2}   & (\bar{\gamma}')^{1/2}
\end{array}\right).
\end{equation}
The connection 
\begin{equation}\label{alpha'}
\bar{\alpha}=\left(\begin{array}{cc}
0 & 1 \\
-{1\over 2}\bar{T} & 0
\end{array}\right)d\bar{z}
\end{equation}
is a connection on $\bar{E}$:
\begin{equation}
\bar{\eta}_\gamma \gamma^* {\bf\bar{\alpha}} \, \bar{\eta}_\gamma^{-1}
+ \bar{\eta}_\gamma d\bar{\eta}_\gamma^{-1} = {\bf\bar{\alpha}}.
\end{equation}
The connection (\ref{alpha'}) can be represented as
\begin{equation}
\bar{\alpha} = {\bf\bar{m}}\, d {\bf\bar{m}}^{-1},
\end{equation}
where
\begin{equation}
{\bf\bar{m}} = \left(\begin{array}{cc}
\bar{u} & -\bar{v} \\
-\bar{u}_{\bar{z}} & \bar{v}_{\bar{z}}
\end{array}\right).
\end{equation}
\begin{lemma} The matrix $\bf\bar{m}$ satisfies
\begin{equation}
\bar{\eta}_\gamma(z) {\bf\bar{m}}(\gamma z) = {\bf\bar{m}}(z) 
\bar{\chi}^*_\gamma.
\end{equation}
\end{lemma}

\newcommand{\hep}[1]{{\tt hep-th/{#1}}}
\newcommand{\gr}[1]{{\tt gr-qc/{#1}}}

\end{document}